\documentclass[12pt,leqno]{article}
\usepackage{amsmath,amssymb,amsthm,amscd}
\newcommand{\C}{\mathbb{C}}
\newcommand{\Z}{\mathbb{Z}}

\newcommand{\N}{\mathbb{N}}

\newcommand{\vac}{|I\,\rangle}

\newcommand{\End}{\operatorname{End}}
\newcommand{\Resx}{\underset{x=0}{\operatorname{Res}}\,}
\newcommand{\Resy}{\underset{y=0}{\operatorname{Res}}\,}
\newcommand{\Resz}{\underset{z=0}{\operatorname{Res}}\,}

\newcommand{\Ker}{\operatorname{Ker}}

\newcommand{\id}{\operatorname{id}}

\newcommand{\mathO}{\mathcal{O}}
\newcommand{\calo}{\mathO}

\newcommand{\rest}[2]{\left.#1\right|_{#2}}

\newcommand{\limit}[2]{\underset{#1\rightarrow#2}{\lim}}
\newcommand{\calv}{\mathcal{V}}
\newcommand{\cals}{\mathcal{S}}
\newcommand{\scalv}{s|_{\calv}}
\newcommand{\pairing}[2]{\langle#1,#2\rangle}
\newcommand{\ket}[1]{|#1\rangle}
\newcommand{\NO}{\,{\raise0.25em\hbox{$\mathop{\hphantom{\cdot}}%
\limits^{_{\circ}}_{^{\circ}}$}}\,}
\theoremstyle{plain}
 \newtheorem{theorem}{Theorem}[subsection]
	\newtheorem{corollary}[theorem]{Corollary}
	\newtheorem{lemma}[theorem]{Lemma}
	\newtheorem{proposition}[theorem]{Proposition}
	
\theoremstyle{remark}
 \newtheorem{definition}[theorem]{Definition}
 \newtheorem{remark}[theorem]{Remark}
 \newtheorem{note}[theorem]{Note}

\begin{document}
\begin{center}
\begin{Large}
On axioms for a vertex algebra\\ and \\
\vskip 6pt
the locality of quantum fields\footnote{%
This work was partially supported by
Grant-in-Aid for Scientific Research,  the
Ministry of Education, Science and Culture.}
\end{Large}
\vskip 3cm
\begin{large}
Atsushi MATSUO
\end{large}
\vskip 0.5cm
Graduate School of Mathematical Sciences,\\
The University of Tokyo,\\
Komaba, Tokyo 153, Japan\\
\vskip 1cm
\begin{large}
Kiyokazu  NAGATOMO
\end{large}
\vskip 0.5cm
Department of Mathematics, \\Graduate School of Science,
Osaka University, \\Toyonaka, Osaka 560, Japan 
\end{center}
\vskip 2cm
\begin{small}
\textbf{Abstract:}
The identities satisfied by two-dimensional
chiral quantum fields are studied from the point
of view of vertex algebras. The Cauchy-Jacobi
identity (or the Borcherds identity) for three
mutually local fields is proved and consequently
a direct proof of Li's theorem on a local
system of vertex operators is provided.
Several characterizations of vertex algebras
are also discussed.

\end{small}
\newpage
\setcounter{section}{-1}
\section{Introduction}
The aim of this paper is to study the identities
satisfied by two-dimensional chiral quantum
fields and to apply them to the basis of the
theory of vertex algebras.

The notion of a vertex algebra was introduced by
Borcherds in \cite{B86} as a purely
algebraic structure: A vertex algebra is a
vector space equipped with a series of
binary operations subject to certain
axioms (see Section 3 for the precise
definition). In the famous work of
Frenkel-Lepowsky-Meurman \cite{FLM}, some of
the properties of a vertex algebra are expressed
in a unified manner: In our notation,
\[
\sum_{i=0}^\infty
\binom{p}{i}(a_{(r+i)}b)_{(p+q -i)}c 
=\sum_{i=0}^\infty (-1)^i\binom{r}{i}\biggl(
a_{(p+r-i)}(b_{(q + i)}c) -
(-1)^rb_{(q+r-i)}(a_{(p+i)}c)\biggr).
\]
This is usually written in terms of the
generating series involving the delta function,
and is called the Cauchy-Jacobi
identity\footnote{We call it the Borcherds
identity since an important case is given in
\cite{B86}.}.

On the other hand, the axioms are related to
some aspects of two-dimensional quantum field
theory (\cite{BPZ},
\cite{G}), and there are vast number of studies
on such structures both in mathematics and in
physics. In this context, the most important
feature of quantum field theory is the
locality of fields (see Subsection 1.5).

Now, in a fundamental work \cite{Li}, Li showed
that a vector space consisting of
pairwise mutually local fields containing the
identity field has a structure of vertex
algebra with respect to the binary operations
defined by
\[
A(z)_{(n)}B(z) = \Resy
A(y)B(z)(y-z)^n|_{|y|>|z|}-
\Resy
B(z)A(y)(y-z)^n|_{|y|<|z|}
\]
if the space is closed under these products.
In \cite{Li}, the Cauchy-Jacobi identity for
these products is shown indirectly as a
consequence of the equivalence of the axioms for
a vertex algebra with the characterization
considered by Goddard in \cite{G}.

In this paper, we shall prove the
identity
\[
\begin{split}
&\sum_{i=0}^\infty
\binom{p}{i}[[A(x),B(y)],C(z)](x-y)^{r+i}(y-z)^{p+q-i}\\
&= \sum_{i=0}^\infty(-1)^i
\binom{r}{i}[A(x),[B(y),C(z)]](x-y)^{p+r-i}(y-z)^{q+i}\\
&\phantom{\sum_{i=0}^\infty(-1)^i
\binom{r}{i}}-\sum_{i=0}^\infty(-1)^{r+i}
\binom{r}{i}[B(y),[A(x),C(z)]](y-z)^{q+r-i}(x-z)^{p+i}.
\end{split}
\]
for mutually local fields $A(z), B(z)$ and $C(z)$
(Theorem \ref{theorem:n1}). Then by taking
the residue of the both sides, we obtain the
Cauchy-Jacobi identity, thus we shall provide an
alternative proof of Li's theorem mentioned above.
Our proof of the identity is purely
algebraic involving careful manipulation of
formal Laurent series and clever use of the
locality. We shall also discuss the proof, under
some additional assumption, by the argument of
contour deformation (see Subsection 6.2).

Another task of the present paper is to give 
transparent proofs of the equivalence of
various characterizations of vertex algebras.
They are: Lian-Zuckerman's characterization by
the locality and bijectivity of the state-field
correspondence (Theorem \ref{theorem:4.5}),
Goddard's characterization mentioned above by
the locality, the creativity\footnote{Our usage
of this term differs from \cite{LZ2}.} and the
translation covariance (Theorem
\ref{theorem:5.4}), and finally the
characterization by the locality, surjectivity
of the state-field correspondence and the
translation covariance (Theorem
\ref{theorem:5.7}). In our treatment, the roles
played by the properties in either
characterization will be clearly recognized.

The paper is organized as follows.
In the first section, we prepare various
notions concerning two-dimensional quantum
field theory in an abstract way using the
language of formal Laurent series.
In the second section, we state and prove our
main theorem.
In section 3, we give the definition of a
vertex algebra, derive various properties, and
describe Li's theorem.
Section 4 and 5 are devoted to the
characterizations of vertex algebras.
In the last section, we discuss the analytic
method of manipulating fields by means of
``operator valued rational function".

For simplicity, we only deal with the bosonic
cases (vertex algebras) in this
paper. However, generalization to the cases
involving both bosonic and fermionic fields
(vertex superalgebras) is straightforward.

\section{Mutually local fields}\label{section:1}
In this section, we describe the definitions and
basic properties concerning two-dimensional
chiral quantum fields in the language of formal
Laurent series.

\subsection{Preliminaries}\label{subsection:1.1}
Throughout
the paper, we always work over a field
$\mathbf{k}$ of characteristic zero. We denote by
$V[[z,z^{-1}]]$ the set of all formal Laurent
series in the variable $z$ with coefficients in a
vector space $V$ possibly having infinitely
many terms both of positive and of negative
degree:
\[
V[[z,z^{-1}]] =\left\{\sum_{n=-\infty}^{\infty}
v_nz^{-n-1}\,\bigg|\,v_n\in V\right\}.
\]
The subset consisting of all series with only
finitely many terms of negative degree is
denoted by
\[
V((z)) = \left\{\sum_{n=n_0}^{\infty}
v_nz^{-n-1}\,\bigg|\,v_n\in V,n_0\in\Z\right\}.
\]
Similarly we write
\begin{align*}
&V[[y,y^{-1},z,z^{-1}]]
=\left\{\sum_{m,n=-\infty}^{\infty}
v_{m,n}y^{-m-1}z^{-n-1}\,\bigg|\,v_{m,n}\in
V\right\},\\
&V((y,z))
=\left\{\sum_{m=m_0}^\infty\sum_{n=n_0}^{\infty}
v_{m,n}y^{-m-1}z^{-n-1}\,\bigg|\,v_{m,n}\in V,
m_0,n_0\in\Z\right\} .
\end{align*}
Let the symbols $(y-z)^n|_{|y|>|z|}$ and
$(y-z)^n|_{|y|<|z|}, (n\in \Z)$, denote the
elements of $\textbf{k}[[y,y^{-1},z,z^{-1}]]$
obtained by expanding the rational function
$(y-z)^n$ into series convergent in the
regions
$|y|>|z|$ and $|y|<|z|$ respectively:
\begin{align*}
(y-z)^n|_{|y|>|z|}&=
\sum_{i=0}^\infty(-1)^i\binom{n}{i}y^{n-i}z^i,\\
(y-z)^n|_{|y|<|z|}&=
\sum_{i=0}^\infty(-1)^{n+i}\binom{n}{i}y^{i}z^{n-i}.
\end{align*}
We will often neglect writing the region of
expansion in case it is obvious from the context
(see Note \ref{note:1.2.3}).

The following lemma will be frequently used later.
\begin{lemma}\label{lemma:1.1.1}
Let $a(y,z)$ be a series with only finitely many terms
of negative or positive degree in $y$ or $z$. If
$a(y,z)$ satisfies $(y-z)^m a(y,z) = 0$ for some
nonnegative integer $m$, then $a(y,z) = 0$.
\end{lemma}
\begin{proof}
Consider the series $b(y,z) = (y-z)^{m-1}a(y,z)$ and
let the coefficients $b_{k,\ell}\in V$ 
be defined 
$b(y,z) =
\sum_{k,\ell\in\Z}b_{k,\ell}y^{-k-1}z^{-\ell
-1}$.
It follows from 
$(y-z)b(y,z) =(y-z)^na(y,z) =0$ that
\begin{equation}\label{eqn:1.1}
b_{k+1,\ell} = b_{k,\ell+1}\quad\text{for all}\quad
k,\ell\in\Z.
\end{equation}
Since $a(y,z)$ has only finitely many terms of negative
or positive degree in $y$ or $z$, so does $b(y,z)$. Thus
(\ref{eqn:1.1}) implies $b(y,z) = 0$. Repeating this
procedure, we arrive at $a(y,z) = 0$.
\end{proof}

\subsection{Series on a vector space}
Let $M$ be a vector space and consider the space
$(\End M)[[z,z^{-1}]]$ of all formal Laurent
series with coefficients being endomorphisms (operators) on
$M$. We simply call such series\footnote{It is called a formal distribution in 
\cite{K96}.}  $A(z)$ a
\textit{series} on $M$. For a
series
$A(z)$ on
$M$, we set
\[
A_n = \Resz A(z)z^n =\text{ coefficient of }
z^{-n-1} \text{ in } A(z)
\]
so that the expansion of $A(z)$ is
\[
A(z) = \sum_{n\in\Z}A_n z^{-n-1},\quad A_n\in
\End M.
\]
The $A_n$ is called a \textit{Fourier mode} 
of $A(z)$. We write
\[
A(z)v = \sum_{n\in\Z}A_nv z^{-n-1}
\]
for a vector $v\in M$. Given two series $A(z)$
and $B(z)$ on $M$, we set
\[
A(y)B(z) =\sum_{m,n\in\Z}A_mB_ny^{-m-1}z^{-n-1},
\]
where $A_mB_n$ denotes
the composition of endomorphisms.

The derivative of a series $A(z)$ is defined by
\[
\partial A(z) = \sum_{n\in\Z}(-n-1)A_nz^{-n-2}
=-\sum_{n\in\Z}nA_{n-1}z^{-n-1}.
\]
More generally, for a nonnegative integer $k\in
\N$, we define
\[
\partial^{(k)}A(z) =
\sum_{n\in\Z}\binom{-n-1}{k}A_nz^{-n-k-1}
=\sum_{n\in\Z}(-1)^k\binom{n}{k}A_{n-k}z^{-n-1}.
\]
Finally, we define the annihilation and creation
parts of a series $A(z)$ by
\[
A(z)_+ = \sum_{n\geq 0}A_{n}z^{-n-1}
\quad \text{and}\quad A(z)_- =
\sum_{n< 0}A_{n}z^{-n-1}
\]
respectively. Then we have 
$\partial (A(z)_\pm) 
= (\partial A(z))_\pm$.

\subsection{Fields on a vector space}
Next we consider the notion of two-dimensional
chiral\footnote{Here \textit{chiral\,} means the
``holomorphic part" in the sense of conformal
field theory.} quantum fields.

\begin{definition}\label{definition:1}
A series $A(z)\in (\End M)[[z,z^{-1}]]$ is called
a \textit{field}\,\footnote{We
followed the terminology in \cite{K96}.} on $M$ if $A(z)v\in M((z))$
for any $v\in M$.
\end{definition}
In other words, $A(z)$ is a field if and only if,
for any $v\in M$, there exists an integer $n_0$
depending on $v$ such that $A_nv= 0$ holds for
all
$n\geq n_0$. Note that if $A(z)$ is a field,
then so is
$\partial A(z)$.

The following simplest example of a field will play
an important role later:
\begin{definition}\label{definition:2}
The \textit{identity field} on $M$ is the field
\[
I(z) = \id_M
\]
of which the only nonnegative term is the
constant term being the identity operator on $M$.
\end{definition}

Now, let $A(z)$ be a series and $B(z)$ a field on
$M$. Consider the following expression
\begin{equation}\label{eqn:1}
\Resy A(y)B(z)(y-z)^m|_{|y|>|z|}
 = \sum_{n\in\Z}\left(
\sum_{i=0}^\infty (-1)^i\binom{m}{i}A_{m-i}B_{n+i}
\right) z^{-n-1}
\end{equation}
for a fixed integer $m$.
Then, for any $v\in M$, the second sum in the
right applied to $v$ is a finite sum for each
$n$ since finitely many $B_{n+i}v$ are nonzero.
Moreover 
$B_{n+i}v$ vanish for sufficiently large $n$.
Therefore (\ref{eqn:1}) defines a field on $M$.

Next, assume further that $A(z)$ is a field,
and consider 
\begin{equation}\label{eqn:2}
\Resy B(z)A(y)(y-z)^m|_{|y|<|z|}=
\sum_{n\in\Z}\left(
\sum_{i=0}^\infty
(-1)^{m+i}\binom{m}{i}B_{m+n-i}A_{i}
\right) z^{-n-1}
\end{equation}
for a fixed integer $m$. Then, for any $v\in M$,
the second sum in the right applied to $v$
is a finite sum for each $n$ since finitely
many
$A_iv$ are nonzero.
Let $A_0v,\dots, A_\ell v$ be the nonzero
vectors. Then, for each $0\leq i\leq \ell$, we have
$B_{m+n_i-i}A_iv = 0$ for sufficiently large $n_i$.
Therefore, taking the maximum of $n_i, (0\leq
i\leq \ell)$, we see that the above summation is
zero for sufficiently large $n$. Hence
(\ref{eqn:2}) gives rise to a field on $M$.

\begin{note}\label{note:1.2.3}
To simplify the presentation of the paper, we often omit the region
of the expansion of $(y-z)^m$ if there is no danger of confusion;
The region is determined by the order of $A(y)$ and $B(z)$. Namely,
we always regard
\[
A(y)B(z)(y-z)^m\quad\text{and}\quad 
B(z)A(y)(y-z)^m
\]
as the series obtained by expanding $(y-z)^m$ convergent in the
regions
\[
|y|>|z|\quad\text{and}\quad 
|y|<|z|
\]
respectively. We also obey this rule in case
more than two series are involved.
\end{note}

Finally, we note that 
\begin{equation}\label{eqn:3}
\begin{split}
&\Resy \partial A(y)B(z)(y-z)^m= -m \Resy
A(y)B(z)(y-z)^{m-1}\\ & = \partial\left(
\Resy A(y)B(z)(y-z)^m\right)
-\Resy A(y)\partial B(z)(y-z)^m
\end{split}
\end{equation}
if $B(z)$ is a field, and
\begin{equation}\label{eqn:4}
\begin{split}
&\Resy B(z)\partial A(y)(y-z)^m= -m \Resy
B(z)A(y)(y-z)^{m-1}\\ & = \partial\left(
\Resy B(z)A(y)(y-z)^m\right)
-\Resy \partial B(z)A(y)(y-z)^m
\end{split}
\end{equation}
if $A(z)$ and $B(z)$ are fields.

\subsection{Residual products of fields}
Now let us explain the residual
products\footnote{In the literature, it is
simply called the $n$-th product. However, to
distinguish it from  the abstract products of a
vertex algebra, we have added the adjective 
\textit{residual}.}\, (\cite{BG}, [Li, Lemma
3.1.4], [LZ1, Definition 2.1]), indexed by
integers, which assign a field to an ordered
pair of fields for each integer
$n$.

\begin{definition}\label{definition:3}
For two fields $A(z)$ and $B(z)$ on $M$, the residual
$m$-th product, $(m\in\Z)$, is defined by
\begin{equation}\label{eqn:5}
A(z)_{(m)}B(z) = \Resy A(y)B(z)(y-z)^m - \Resy B(z)A(y)(y-z)^m.
\end{equation}
\end{definition}
Explicitly, (\ref{eqn:5}) is written as
\[
A(z)_{(m)}B(z) =\sum_{n\in\Z}
(A_{(m)}B)_n
z^{-n-1},
\]
where
\[
(A_{(m)}B)_n =\sum_{i=0}^\infty
(-1)^i\binom{m}{i}(A_{m-i}B_{n+i}
-(-1)^mB_{m+n-i}A_i)
.
\]
In the physics notation, the residual products
are expressed as (cf. [BG, (2.11)]):
\[
\begin{split}
&A(z)_{(n)}B(z)=\oint_{C_z}\frac{dy}{2\pi\sqrt{-1}}
R(A(y)B(z))(y-z)^n\\
&=\oint_{|y|>|z|}\frac{dy}{2\pi\sqrt{-1}}A(y)B(z)(y-z)^n
-
\oint_{|y|<|z|}\frac{dy}{2\pi\sqrt{-1}}B(z)
A(y)(y-z)^n
\end{split}
\]
where $R$ denotes the \textit{radial ordering}\,
defined by
\begin{align*}
R(A(y)B(z)) = 
\begin{cases}
A(y)B(z),&\quad |y|>|z|,\\
B(z)A(y),&\quad |y|<|z|,
\end{cases}
\end{align*}
and $C_z$ is a small contour around
$z$.

Now, since $A(z)$ and $B(z)$ are fields,
$A(z)_{(m)}B(z)$ is again a field on $M$ by the
consideration in the preceding subsection.

\begin{definition}\label{definition:1.4.3}
The space $\langle\mathcal{S}\,\rangle$
\textit{generated by}\, a set of fields
$\mathcal{S}$ is the linear span of the fields
constructed by successive application of the
residual products to the fields in
$\mathcal{S}$ as well as the identity field
$I(z)$.
\end{definition}

The space $\langle\mathcal{S}\,\rangle$ is the
smallest space closed under the residual
products that contains $\mathcal{S}\cup
\{I(z)\}$.

\medskip
We always understand that the derivative
proceeds the residual product:
\[
\partial A(z)_{(m)}B(z) = (\partial
A(z))_{(m)}B(z),\quad A(z)_{(m)}\partial B(z) =
A(z)_{(m)}(\partial B(z)).
\]
Then, by (\ref{eqn:3}) and (\ref{eqn:4}), 
we have
\begin{equation}\label{eqn:6}
\partial A(z)_{(m)}B(z)= -m A(z)_{(m-1)}B(z)
=\partial (A(z)_{(m)}B(z)) - A(z)_{(m)}\partial
B(z).
\end{equation}
In particular, let us consider the $(-1)$ st
product:
\[
A(z)_{(-1)}B(z) =\sum_{n\in\Z}
\left(
\sum_{i=0}^\infty (A_{-i-1}B_{n+i}
+B_{n-i-1}A_i)
\right)
z^{-n-1}.
\]
It coincides with what is called the \textit{normally ordered
product}\,\footnote{cf. [BBS, (A.2)], [SY, (5.5)]}:
\[
\NO A(z)B(z)\NO= A(z)_-B(z) + B(z)A(z)_+.
\]
Therefore, using (\ref{eqn:6}), we have
\[
A(z)_{(-k-1)}B(z) = \NO \partial^{(k)}A(z)B(z)\NO
\]
for a nonnegative integer $k$. Note that 
$\NO A(z)B(z)\NO \neq \NO B(z)A(z)\NO$ in
general.

Let us close this subsection with the following:
\begin{proposition}\label{proposition:1}
Let $A(z)$ be a field and $I(z)$ the identity field on a vector
space $M$. Then 
\begin{equation}\label{eqn:7}
A(z)_{(m)}I(z)=
\begin{cases}
0,&\quad (m\geq 0),\\
\partial^{(-m-1)}A(z),&\quad (m\leq -1),
\end{cases}
\end{equation}
\begin{equation}\label{eqn:8}
I(z)_{(m)}A(z)=
\begin{cases}
0,&\quad (m\neq -1),\\
A(z),&\quad (m= -1).
\end{cases}
\end{equation}
\end{proposition}
We note, in particular,
\begin{equation}\label{eqn:9}
A(z)_{(m)}I(z)=
\begin{cases}
0,&\quad (m\geq 0),\\
A(z),&\quad (m = -1),
\end{cases}
\end{equation}
which is a part of the axioms 
for a vertex algebra (see Subsection 3.1).

\subsection{Locality of fields}
The notion of locality for two-dimensional 
chiral quantum fields
is related to Wightman's axioms for quantum field
theory (cf.[K, Chapter 1]). We adopt the
following formulation in the language of formal
Laurent series%
\footnote{
The condition (\ref{eqn:10}) was first considered by
Dong-Lepowsky [DL, (7.24)] under the term
\textit{commutativity}, while the same was
considered in earlier papers in the language of
operator valued rational functions (cf.
[G],[FLM],[FHL]). } :

\begin{definition}\label{definition:4}
Two fields $A(z)$ and $B(z)$ are called (mutually) \textit{local}
if
\begin{equation}\label{eqn:10}
A(y)B(z)(y-z)^n = B(z)A(y)(y-z)^n
\end{equation}
holds for some nonnegative integer $n$. In
this case, we also say
$A(z)$ is local to $B(z)$.
\end{definition}

In terms of Fourier modes, (\ref{eqn:10}) is written as
\[
\sum_{i=0}^n (-1)^i\binom{n}{i}A_{p+n-i}B_{q+i} =
\sum_{i=0}^n (-1)^i\binom{n}{i}B_{q+i}A_{p+n-i},
\]
or equivalently
\[
\sum_{i=0}^n (-1)^i\binom{n}{i}\left(
A_{p+n-i}B_{q+i} -(-1)^nB_{q+n-i}A_{p+i}
\right) =0
\]
where $p,q$ run over all integers. Note that $A(z)$ need not be
local to itself.

Let us introduce the following notion:

\begin{definition}\label{definition:5}
The \textit{order} of locality of fields $A(z)$ and $B(z)$ is the
minimum of the nonnegative integers $n$
satisfying (\ref{eqn:10}).
\end{definition}
Thus, $A(z)$ and $B(z)$ are local at order $n_0$
if and only if  (\ref{eqn:10}) holds precisely
for $n\geq n_0$.
Then, by the definition of the residual products,
we have
\begin{equation}\label{eqn:1.12}
A(z)_{(n)}B(z) = 0,\quad(n\geq n_0).
\end{equation}
Note that it may happen that $A(z)_{(n)}B(z) =
0$ holds for some $0\leq n<n_0$. It is easy to see
that if $A(z)$ and $B(z)$ are local at order
$n_0\,(\geq 1)$, then $\partial A(z)$ and $B(z)$
are local at order
$n_0+1$.

Now we turn to the study of the locality 
of many fields. We say that fields
$A^{(1)}(z), \dots, A^{(\ell)}(z)$ are local
if all the distinct pairs $A^{(i)}(z)$ and
$A^{(j)}(z), (i<j)$, are local.

The following proposition is a generalization
of [Li, proof of Proposition 3.2.7]. Recall our
convention in Note \ref{note:1.2.3}

\begin{proposition}\label{proposition:1.5.6}
Let $A^{(1)}(z),\dots, A^{(\ell)}(z)$ be local
fields and let $m_{ij}$ be the order of
locality of
$A^{(i)}(z)$ and $A^{(j)}(z)$ for each $i<j$.
Then
\[
[\cdots
[[A^{(1)}(z_1),A^{(2)}(z_2)],A^{(3)}(z_3)],
\cdots,A^{(\ell)}(z_\ell)]
\smash{\prod_{i<j}}(z_i-z_j)^{n_{ij}} = 0
\]
holds if $\sum_{i<j}n_{ij}\geq \sum_{i<j}m_{ij}
-(\ell-2)$.
\end{proposition}
\begin{proof}
We show by induction on $\ell$. The case $\ell = 2$
is nothing but the definition of the locality.
Suppose $\ell>2$. Then, by expanding as
\[
\begin{split}
(z_i-z_\ell)^{n_{i\ell}}&=
(z_i-z_j+z_j-z_\ell)^{n_{i\ell}-m_{i\ell}}
(z_i-z_\ell)^{m_{i\ell}}\\
&= \sum_{s=0}^{n_{i\ell}-m_{i\ell}}
\binom{n_{i\ell}-m_{i\ell}}{s}
(z_i-z_j)^{n_{i\ell}-m_{i\ell}-s}
(z_j-z_\ell)^{s}
(z_j-z_\ell)^{m_{i\ell}}
\end{split}
\]
for an appropriate $j$ if $n_{i\ell}\geq m_{i\ell}$,
and by repeating this procedure, the left-hand
side of the desired equality is written as a
linear combination of terms of the form
\begin{equation}\label{eqn:1.5.13}
[\cdots
[[A^{(1)}(z_1),A^{(2)}(z_2)],A^{(3)}(z_3)],
\cdots,A^{(\ell)}(z_\ell)]
\smash{\prod_{i<j}}(z_i-z_j)^{p_{ij}}
\end{equation}
where the exponents $p_{ij}$ satisfy
\[
\sum_{1\leq i<j\leq \ell-1}p_{ij}\geq \sum_{1\leq
i<j\leq \ell-1}m_{ij} - (\ell-3), \quad
\text{or} 
\quad p_{i\ell}\geq m_{i\ell} \text{ for
all } i.
\]
In the former case, (\ref{eqn:1.5.13}) vanishes
by the inductive hypothesis, while in the latter
case, by the locality of $A^{(i)}(z)$ and
$A^{(\ell)}(z)$. 
\end{proof}

In particular, we have
\begin{lemma}\label{lemma:1.2}
Let $A(z), B(z)$ and $C(z)$ be local fields and let
$k_0,\ell_0$ and
$m_0$ be the order of locality of $A(z)$ and
$C(z)$, $B(z)$ and
$C(z)$, and $A(z)$ and $B(z)$ respectively. 
Then, for any integers 
$k,\ell,m$,
\begin{multline*}
(y-z)^n\Bigl(
A(x)B(y)C(z) - B(y)A(x)C(z)\Bigr)(x-y)^m (y-z)^\ell
(x-z)^k
\\ =
(y-z)^n\Bigl(
C(z)A(x)B(y) - C(z)B(y)A(x)\Bigr)(x-y)^m (y-z)^\ell
(x-z)^k
\end{multline*}
holds for all $n\in \N$ satisfying $n\geq k_0
+\ell_0 +m_0 -k-\ell- m-1$.
\end{lemma}
An immediate consequence of this is ([Li,
Proposition 3.2.7])
\begin{proposition}\label{proposition:2}
If $A(z), B(z)$ and $C(z)$ are local, then
$A(z)_{(m)}B(z)$ and $C(z)$ are local.
\end{proposition}

Here the order of locality of
$A(z)_{(m)}B(z)$ and
$C(z)$ is at most $k_0+\ell_0+m_0 -m -1$ for
$m<m_0$ whereas $A(z)_{(m)}B(z)= 0$ for $m\geq
m_0$.

Another consequence of the locality is the
following:

\begin{proposition}\label{proposition:1.5.7}
Let $A^{(1)}(z),\dots, A^{(\ell)}(z)$ be local
fields. Then, for any $u\in M$,
\begin{equation}\label{eqn:1.5.14}
A^{(1)}_{p_1}\cdots A^{(\ell)}_{p_\ell}u = 0,
\quad (p_1+\cdots +p_\ell\geq
n,p_1,\dots,p_\ell\in \Z),
\end{equation}
for sufficiently large $n$.
\end{proposition}
\begin{proof}
Since $A^{(i)}(z)$ are fields, we have
$A^{(i)}_pu = 0, (p\geq n_i)$,
for sufficiently large $n_i$. 
Let $m_{ij}$ be the order of locality of
$A^{(i)}(z)$ and $A^{(i)}(z)$ and set
$
n = \sum_{i=1}^\ell n_i + \sum_{1\leq i<j\leq
\ell}m_{ij} -\ell+1.
$
We shall prove (\ref{eqn:1.5.14}) for this $n$
by induction on
$\ell$. It trivially holds in the case $\ell=1$.
We suppose
$\ell>1$. Then, by the locality, we have
\begin{multline*}
A^{(1)}(z_1)\cdots A^{(\ell-1)}(z_{\ell-1})
A^{(\ell)}(z_{\ell})
\smash{\prod_{i=1}^{\ell-1}}(z_i-z_\ell)^{m_{i\ell}}\\
=A^{(\ell)}(z_{\ell}) A^{(1)}(z_1)\cdots
A^{(\ell-1)}(z_{\ell-1})
\smash{\prod_{i=1}^{\ell-1}}(z_i-z_\ell)^{m_{i\ell}}.
\end{multline*}
Therefore, we have
\begin{equation}\label{eqn:1.5.15}
\begin{split}
&\sum_{k_1,\dots,k_{\ell-1}\geq 0}
\prod_{i=1}^{\ell-1}
(-1)^{k_i}
\binom{m_{i\ell}}{k_i}
A^{(1)}_{p_1-k_1}\cdots 
A^{(\ell-1)}_{p_{\ell-1}-k_{\ell-1}}
A^{(\ell)}_{p_\ell+k_1+\cdots +k_{\ell-1}}u\\
& = \sum_{k_1,\dots,k_{\ell-1}\geq 0}
\prod_{i=1}^{\ell-1}
(-1)^{k_i}
\binom{m_{i\ell}}{k_i}
A^{(\ell)}_{p_\ell +\sum_{i=1}^{\ell-1}(m_{i\ell}
-k_{i})}
A^{(1)}_{p_1-m_{i\ell}+k_1}\cdots 
A^{(\ell-1)}_{p_{\ell-1}-m_{\ell-1}+k_{\ell-1}}u
\end{split}
\end{equation}
for any $p_1,\dots,p_\ell\in\Z$.

Now, suppose $p_1+\cdots+p_\ell\geq n$. If
$p_\ell\geq n_\ell$, then
$A^{(1)}_{p_1}\cdots A^{(\ell-1)}_{p_{\ell-1}}
A^{(\ell)}_{p_{\ell}}u = 0$.
If $p_\ell<n_\ell$, then we have
\[
\sum_{i=1}^{\ell-1}(p_i-m_{i\ell})\geq
\sum_{i=1}^{\ell-1}n_i +
\sum_{1\leq i<j\leq \ell-1}m_{ij} -\ell +2
\]
so that the right-hand side of
(\ref{eqn:1.5.15}) vanishes by the induction
assumption. Hence we have
\begin{multline*}
\sum_{k_1,\dots,k_{\ell-1}\in\N}
(-1)^{k_1+\cdots +k_{\ell-1}}
\binom{m_{1\ell}}{k_1}\cdots
\binom{m_{\ell-1\ell}}{k_{\ell-1}}
A^{(1)}_{p_1-k_1}\cdots 
A^{(\ell-1)}_{p_{\ell-1}-k_{\ell-1}}
A^{(\ell)}_{p_\ell+k_1+\cdots
+k_{\ell-1}}u=0.
\end{multline*}
Therefore, by induction on $n_\ell-p_\ell$ for
fixed
$p_1+\cdots+p_\ell(\geq n)$, we have
(\ref{eqn:1.5.14}).
\end{proof}

Finally let us consider a set $\cals$ of
fields. We say that $\cals$ is
\textit{pairwise local}\, if any pair of fields
in $\cals$, not necessarily distinct, are
local. Consider the space $\langle
\cals\,\rangle$ generated by $\cals$. By
successive use of Lemma
\ref{lemma:1.2}, we have

\begin{proposition}\label{proposition:1.5.71}
If a set $\cals$ of fields is pairwise local,
then so is the space $\langle\cals\,\rangle$.
\end{proposition}

\subsection{Operator product
expansion}\label{subsection:1.5} This subsection is
devoted to the explanation of the notion of
operator product expansion from the point of view
of Kac [K, p.20], however, we reformulate it so
that we do not use the delta function.
The results of this subsection will not be used
in the rest of the paper.

Let $A(z)$ and $B(z)$ be local fields 
on a vector space $M$. Then 
$[A(y),B(z)](y-z)^m = 0$
for some $m \in \N$ and we have
\[
[A(y)_+,B(z)] (y-z)^m = -[A(y)_-,B(z)](y-z)^m.
\]
Since the left-hand sides does not have terms
of degree greater than
$m-1$ in $y$ whereas the right does not have
terms of negative degree in $y$, they are equal
to a polynomial of degree
$m-1$ in $y$. Hence we may write
\begin{align*}
[A(y)_+,B(z)] (y-z)^m& =
\sum_{i=0}^{m-1}C^i(z)(y-z)^{m-i-1},
\\
-[A(y)_-,B(z)] (y-z)^m& =
\sum_{i=0}^{m-1}C^i(z)(y-z)^{m-i-1}
\end{align*}
where $C^i(z)$ are some series.
Now, since the difference
$[A(y)_+,B(z)]-\sum_{i=0}^{m-1}C^i(z)/(y-z)^m|_{|y|>|z|}$
has only finitely many terms of positive degree in
$y$, and it vanishes if we multiply it by $(y-z)^m$,
it must be identically zero by Lemma
\ref{lemma:1.1.1}.
Therefore, we have
\[
[A(x)_+, B(z)] =
\sum_{i=0}^{m-1}
\left.\frac{C^i(z)}{(y-z)^{i+1}}\right|_{|y|>|z|}
\]
and similarly
\[
-[A(y)_-,B(z)] =
\sum_{i=0}^{m-1}\left.\frac{C^i(z)}{(y-z)^{i+1}}\right|_{|y|<|z|}.
\]
Therefore
\begin{align}
A(y)B(z)& =
\sum_{i=0}^{m-1}\left.\frac{C^i(z)}{(y-z)^{i+1}}\right|_{|y|>|z|}+\NO
A(y)B(z)\NO\label{eqn:403},\\
B(z)A(y)& =
\sum_{i=0}^{m-1}\left.\frac{C^i(z)}{(y-z)^{i+1}}\right|_{|y|<|z|}+\NO
A(y)B(z)\NO\label{eqn:404}
\end{align}
where $\NO A(y)B(z)\NO = A(y)_-B(z) + B(z)A(y)_+$.

Conversely, if (\ref{eqn:403}) and (\ref{eqn:404})
hold , then it is obvious that
the fields
$A(z)$ and
$B(z)$ are local.

Now, it follows from (\ref{eqn:403}) and (\ref{eqn:404}),
that
$A(z)_{(j)}B(z) = C^j(z)$
by the definition of the residual products.

Hence we have obtained ([K,Theorem 2.3])
\begin{theorem}[Operator product expansion]
\label{theorem:ope}
Let $A(z)$ and $B(z)$ be fields on a vector space. 
They are local 
if and only if both
\begin{align*}
A(y)B(z) &=
\sum_{i=0}^{m-1}\left.\frac{A(z)_{(i)}B(z)}{(y-z)^{i+1}}\right|_{|y|>|z|}+\NO
A(y)B(z)\NO,\\
B(z)A(y) &=
\sum_{i=0}^{m-1}\left.\frac{A(z)_{(i)}B(z)}{(y-z)^{i+1}}\right|_{|y|<|z|}+\NO
A(y)B(z)\NO
\end{align*}
hold for some $m\in\N$.
\end{theorem}

These two equalities in the theorem are often
abbreviated into the single expression
\[
A(y)B(z)\sim
\sum_{i=0}^{m-1}\frac{A(z)_{(i)}B(z)}{(y-z)^{i+1}},
\]
which is called 
the \textit{operator product expansion} (OPE),
and the right-hand side is called the
\textit{contraction}.

\begin{remark}\label{remark:1.5}
The first equality of Theorem \ref{theorem:ope}
holds without the assumption of the locality
([LZ1, Proposition 2.3]):
\[
A(y)B(z) = \sum_{i=0}^\infty
\rest{\frac{A(z)_{(i)}B(z)}{(y-z)^{i+1}}}{|y|>|z|} +
\NO A(y)B(z)\NO,
\]
where the sum in the right  is indeed a
finite sum for each degree in $y$.
\end{remark}

Next let us further expand the
remainder $\NO A(y)B(z)\NO$.
To this end, we prepare the notion of a field in two
variables: A series $A(y,z) =
\sum_{p,q\in\Z}A_{p,q}y^{-p-1}z^{-q-1}$ 
is a field
if, for any $u\in M$, there exists integers 
$p_0$ and $q_0$ such that
\[
A_{p,q}u = 0,\quad \text{if}\quad p\geq
p_0\quad\text{or}\quad q\geq q_0.
\]
In other words, $A(y,z)$ is a field if and
only if $A(y,z)u\in M((y,z))$.

If $A(z)$ and $B(z)$ are fields, then the normally
ordered product $\NO A(y)B(z)\NO$ is a field.
If $A(y,z)$ is a field, then 
\[
A(z,z) = \sum_{p,q\in\Z}A_{p,q}z^{-p-q-2}
=\sum_{n\in\Z}\biggl(\sum_{p+q =
n-1}A_{p,q}\biggr)z^{-n-1}
\]
makes sense and is a field.

\begin{lemma}\label{lemma:1.6}
If $A(y,z)$ is a field, then there exists a unique
field $R(y,z)$ such that
\[
A(y,z) - A(z,z) = (y-z)R(y,z).
\]
\end{lemma}
In fact, the series
\[
R(y,z) = \sum_{p\leq -1,q\in\Z}\biggl(
\sum_{i=0}^\infty
A_{p-i,q+i}\biggr)y^{-p-1}z^{-q-1}
-\sum_{p\geq 0,q\in\Z}
\biggl(\sum_{i=0}^\infty A_{p+i,q-i-1}\biggr)
y^{-p-1}z^{-q-1}
\]
is a field satisfying $A(y,z) - A(z,z)=
(y-z)R(y,z)$. The uniqueness is obvious by Lemma
\ref{lemma:1.1.1}.

By successive use of this lemma, we obtain
([K, Lemma 3.1])

\begin{proposition}[Taylor's
formula]\label{proposition:1.7} If $A(y,z)$ is a
field, then for any positive integer $N$, there
exists a unique field $R_N(y,z)$ such that 
\[
A(y,z) =
\sum_{i=0}^{N-1}\partial^{(i)}_yA(y,z)|_{y=z}(y-z)^i
+ R_N(y,z)(y-z)^N.
\]
\end{proposition}
In particular, for a field $A(z)$, we have
\[
A(y) = \sum_{i=0}^{N-1}\partial^{(i)}_z A(z)(y-z)^i
+ R_N(y,z)(y-z)^N
\]
for some field $R_N(y,z)$.

Therefore ([K,Theorem 3.1])

\begin{theorem}\label{theorem:1.8}
Let $A(z)$ and $B(z)$ be fields on a vector space. 
If they are local at order $m$, then, for any
positive integer $N$, there exists a unique field
$R_N(y,z)$ such that
\begin{align*}
A(y)B(z)& =
\sum_{i=-N}^{m-1}\rest{\frac{A(z)_{(i)}B(z)}{(y-z)^{i+1}}}{|y|>|z|}
+ R_N(y,z)(y-z)^N,\\
B(z)A(y)& =
\sum_{i=-N}^{m-1}\rest{\frac{A(z)_{(i)}B(z)}{(y-z)^{i+1}}}{|y|<|z|}
+ R_N(y,z)(y-z)^N.
\end{align*}
\end{theorem}

This result is seen to be an interpretation 
in the language of formal Laurent series of the
expressions
\[
A(y)B(z) =
\sum_{i=-\infty}^{m-1}\frac{A(z)_{(i)}B(z)}{(y-z)^{-i-1}},
\quad
B(z)A(y)=
\sum_{i=-\infty}^{m-1}\frac{A(z)_{(i)}B(z)}{(y-z)^{-i-1}}.
\]

\section{Borcherds identity for local fields}
\label{section:2}
In this section, we show the identity 
satisfied by three
local fields with respect to the residual products, which is a
consequence of the usual Jacobi identity
\[
[[A(x),B(y)],C(z)] = [A(x),[B(y),C(z)]] - [B(y),[A(x),C(z)]].
\]
The strategy is, roughly speaking, to multiply this Jacobi identity
by the rational function
\[
(x-y)^r(y-z)^q(x-z)^p
\] 
and take the residue after expanding it in various regions. Special
cases of such derivation are considered by Li [Li, p.166] and Kac
[K, Proposition 2.3 (c) and Proposition 3.3 (c)]. However, to execute
it in full generality, one has to be careful about divergence,
which can be avoided by clever use of the locality.

\subsection{Binomial identities}
Let us consider the rational function
\[
F(x,y,z) = (x-y)^r(y-z)^q(x-z)^p,
\quad 
p,q,r \in\Z,
\]
which has the expansions 
\begin{alignat*}{2}
F_0(x,y,z) & = \sum_{i=0}^\infty\binom{p}{i}
(x-y)^{r+i}(y-z)^{p+q-i},&\quad (|y-z|>|x-y|),\\
F_1(x,y,z) & = \sum_{i=0}^\infty(-1)^i\binom{r}{i}
(x-z)^{p+r-i}(y-z)^{q+i},&\quad (|x-z|>|y-z|),\\
F_2(x,y,z) & = \sum_{i=0}^\infty(-1)^{r+i}\binom{r}{i}
(y-z)^{q+r-i}(x-z)^{p+i},&\quad (|y-z|>|x-z|)
\end{alignat*}
convergent in the respective regions.

Expanding these series again into series in $x,y,z$,
and comparing them with the corresponding expansions
of
$F(x,y,z)$, we have, for example,
\[
F(x,y,z)|_{|x|>|y|>|z|} = F_1(x,y,z)|_{|x|>|y|>|z|},
\]
for all integers $p,q$ and $r$.
To be precise, we have
\[
\begin{split}
\sum_{i,j,k\geq 0}
(-1)^{i+j+k}&\binom{r}{i}\binom{q}{j}\binom{p}{k}
x^{p+r-i-k}y^{q+i-j}z^{j+k}\\ &= 
\sum_{i,j,k\geq 0}
(-1)^{i+j+k}\binom{r}{i}\binom{p+r-i}{j}\binom{q+i}{k}
x^{p+r-i-j}y^{q+i-k}z^{j+k}
\end{split}
\]
which is equivalent to 
a set of identities for binomial coefficients
(See Appendix for detail).

Now, let $A(z),B(z)$ and $C(z)$ be fields on a vector space $M$
which are not necessarily local.
The above argument shows that we have the following set of
identities involving these fields:

\begin{lemma}\label{lemma:2.1}
For all $p,q, r\in\Z$, we have
\begin{align*}
A(x)B(y)C(z)F(x,y,z) & = A(x)B(y)C(z)F_1(x,y,z),\\
A(x)C(z)B(y)F(x,y,z) & = A(x)C(z)B(y)F_1(x,y,z),\\
B(y)A(x)C(z)F(x,y,z) & = B(y)A(x)C(z)F_2(x,y,z),\\
B(y)C(z)A(x)F(x,y,z) & = B(y)C(z)A(x)F_2(x,y,z),\\
C(z)A(x)B(y)F(x,y,z) & = C(z)A(x)B(y)F_0(x,y,z),\\
C(z)B(y)A(x)F(x,y,z) & = C(z)B(y)A(x)F_0(x,y,z),
\end{align*}
where $A(z),B(z),C(z)$ are fields on a vector space $M$.
\end{lemma}

Here we have omitted the regions of the expansions according to our
convention in Note \ref{note:1.2.3}.

Note that, for example, the identity
\[
A(x)B(y)C(z)F(x,y,z) = A(x)B(y)C(z)F_0(x,y,z)
\]
is not valid in general. In fact, 
\[
F_0(x,y,z)|_{|x|>|y|>|z|} =\sum_{i,j,k\geq 0}
(-1)^{j+k}\binom{p}{i}\binom{r+i}{j}\binom{p+q-i}{k}
x^{r+i-j}y^{p+q-i+j-k}z^{k}
\]
is divergent for negative $p$ since each of the coefficients becomes
an infinite sum. On the other hand, the coefficients are finite sums
for nonnegative $p$ and the equality is valid
in this case.

Such cases are summarized as follows:
\begin{lemma}\label{lemma:2.2}
For all $p\in\N$ and $p,q\in\Z$, we have
\begin{align*}
A(x)B(y)C(z)F(x,y,z) & = A(x)B(y)C(z)F_0(x,y,z),\\
B(y)A(x)C(z)F(x,y,z) & = B(y)A(x)C(z)F_0(x,y,z).
\end{align*}
For all, $p,q\in\Z$ and all $r\in\N$, we have
\begin{align*}
B(y)C(z)A(x)F(x,y,z) & = B(y)C(z)A(x)F_1(x,y,z),\\
C(z)B(y)A(x)F(x,y,z) & = C(z)B(y)A(x)F_1(x,y,z),\\
A(x)C(z)B(y)F(x,y,z) & = A(x)C(z)B(y)F_2(x,y,z),\\
C(z)A(x)B(y)F(x,y,z) & = C(z)A(x)B(y)F_2(x,y,z).
\end{align*}
\end{lemma}

\subsection{Borcherds identity for non-local
fields} If $p$ and $r$ are both nonnegative,
then all the identities of Lemma \ref{lemma:2.1}
and  Lemma \ref{lemma:2.2} are valid. 
Therefore
\[
\begin{split}
[[A(x),&B(y)],C(z)]F_0(x,y,z)=
[[A(x),B(y)],C(z)]F(x,y,z)\\ 
&=
\left([A(x),[B(y),C(z)]]-[B(y),[A(x),C(z)]]\right)F(x,y,z)\\
&=
[A(x),[B(y),C(z)]]F_1(x,y,z)-[B(y),[A(x),C(z)]]F_2(x,y,z).
\end{split}
\]
Thus we have obtained 
\begin{theorem}\label{theorem:n1}
Let $A(z),B(z)$ and $C(z)$ be fields on a vector space.
Then, for any $p,q\in\N$ and any $q\in\Z$,
\[
\begin{split}
&\sum_{i=0}^\infty
\binom{p}{i}[[A(x),B(y)],C(z)](x-y)^{r+i}(y-z)^{p+q-i}\\
&= \sum_{i=0}^\infty(-1)^i
\binom{r}{i}[A(x),[B(y),C(z)]](x-y)^{p+r-i}(y-z)^{q+i}\\
&\phantom{\sum_{i=0}^\infty(-1)^i
\binom{r}{i}}-\sum_{i=0}^\infty(-1)^{r+i}
\binom{r}{i}[B(y),[A(x),C(z)]](y-z)^{q+r-i}(x-z)^{p+i}.
\end{split}
\]

\end{theorem}
Taking $\Resy\Resx$ of the both sides, we
have\footnote{A special case  $(r = 0)$ of
this result is described by Kac [K, Proposition
3.3 (c)]. General case is deduced from this
case by the inductive structure of the Borcherds
identity (cf. Proposition 3.1).}

\begin{corollary}\label{corollary:2.1}
Let $A(z),B(z)$ and $C(z)$ be fields 
on a vector space. Then, for
any $p,r\in\N$ and any $q\in\Z$,
\[
\begin{split}
&\sum_{i=0}^\infty \binom{p}{i}\left(
A(z)_{(r+i)}B(z)\right)_{(p+q-i)}C(z)\\
&= \sum_{i=0}^\infty
(-1)^i\binom{r}{i}
\left(
A(z)_{(p+r-i)}(B(z)_{(q+i)}C(z))
- (-1)^rB(z)_{(q+r-i)}(A(z)_{(p+i)}C(z))
\right).
\end{split}
\]
\end{corollary}

This result is not true in general if $p$ or $r$ is negative.  It is
because of the failure of Lemma \ref{lemma:2.2} for such indices.
However, if we assume the locality of the fields
$A(z),B(z)$ and $C(z)$, then the theorem is
generalized to arbitrary integers $p,q,r$, as
we will see in the following two sections.

\subsection{Consequences of locality}
Now assume that the fields $A(z), B(z)$ and $C(z)$ are local.
Then, by Lemma \ref{lemma:1.2}, we have
\begin{multline*}
(y-z)^n [A(x),B(y)]C(z)(x-y)^k(y-z)^\ell(x-z)^m\\
=(y-z)^n C(z)[A(x),B(y)](x-y)^k(y-z)^\ell(x-z)^m
\end{multline*}
for sufficiently large $n$. This reduces a calculation involving
the left-hand side to that involving 
the right-hand side.

In this way, we obtain the following lemma.
\begin{lemma}\label{lemma:2.6}
If $A(z), B(z)$ and $C(z)$ are local, then for
any $p,q,r\in\Z$,
\begin{align*}
[A(x),B(y)]C(z)F(x,y,z) &=
\sum_{i=0}^\infty \binom{p}{i}[A(x),B(y)]C(z)
(x-y)^{r+i}(y-z)^{p+q-i},\\
[B(y),C(z)]A(x)F(x,y,z) &=
\sum_{i=0}^\infty (-1)^i\binom{r}{i}[B(y),C(z)]A(x)
(x-z)^{p+r-i}(y-z)^{q+i},\\
[A(x),C(z)]B(y)F(x,y,z) &=
\sum_{i=0}^\infty (-1)^{r+i}\binom{r}{i}[A(x),C(z)]B(y)
(y-z)^{q+r-i}(x-z)^{p+i}.
\end{align*}
Here the right-hand side of each equality is a
finite sum because  of the locality.
\end{lemma}

\begin{proof}
By Lemma \ref{lemma:1.2} and Lemma \ref{lemma:2.2},
\[
\begin{split}
&(y-z)^n\sum_{i=0}^\infty \binom{p}{i}[A(x),B(y)]C(z)
(x-y)^{r+i}(y-z)^{p+q-i}\\
&= (y-z)^n\sum_{i=0}^\infty \binom{p}{i}
C(z)[A(x),B(y)](x-y)^{r+i}(y-z)^{p+q-i}\\
&=(y-z)^nC(z)[A(x),B(y)]F_0(x,y,z)\\
&= (y-z)^nC(z)[A(x),B(y)]F(x,y,z)\\
&= (y-z)^n[A(x),B(y)]C(z)F(x,y,z)
\end{split}
\]
for sufficiently large $n$. Therefore the series
\[
\begin{split}
D(y,z)= \sum_{i=0}^\infty \binom{p}{i}
\biggl([A(x),&B(y)]C(z)(x-y)^{r+i}(y-z)^{p+q-i}\\
&-[A(x),B(y)]C(z)F(x,y,z)\biggr)
\end{split}
\]
satisfies $(y-z)^nD(y,z)v = 0$ for any vector $v\in M$.
Moreover, since
$D(y,z)v$ has only finitely many terms of negative degree in $z$, we
must have
$D(y,z)v = 0$ by Lemma \ref{lemma:1.1.1}. Thus we have
shown the first equality of the lemma. The other
equalities are proved similarly.
\end{proof}

\begin{remark}
Note that the right-hand side of the first
equality, for example, makes sense if $A(z)$ and
$B(z)$ are local.  However, in order the
equality to hold, we need not only this but also
the locality of $A(z)$ and $C(z)$,  and of
$B(z)$ and $C(z)$ in general.
\end{remark}

\subsection{Borcherds identity for local fields}
Now we arrive at the main result of this paper.
\begin{theorem}\label{theorem:n2}
Let $A(z),B(z)$ and $C(z)$ be fields on a
vector space. If they are local, then for any
$p,q,r\in\Z$,
\[
\begin{split}
&\sum_{i=0}^\infty \binom{p}{i}
[[A(x),B(y)],C(z)]
(x-y)^{r+i}(y-z)^{p+q-i}\\
&= \sum_{i=0}^\infty (-1)^{i}\binom{r}{i}\biggl(
[A(x),[B(y),C(z)]](x-z)^{p+r-i}(y-z)^{q+i} \\
&\phantom{\sum_{i=0}^\infty
(-1)^{i}\binom{r}{i}\biggl(
[A(x),}- (-1)^r[B(y),[A(x),C(z)]]
(y-z)^{q+r-i}(x-z)^{p+i}\biggr).
\end{split}
\]
\begin{proof}
By Lemma \ref{lemma:2.1} and Lemma
\ref{lemma:2.6}, we have
\[
\begin{split}
&\sum_{i=0}^\infty\binom{p}{i}\left(
[A(x),B(y)]C(z)- C(z)[A(x),B(y)]\right)
(x-y)^{r+i}(y-z)^{p+q-i}\\
&=\left(
[A(x),B(y)]C(z) - C(z)[A(x),B(y)]\right) F(x,y,z)\\
&= (A(x)[B(y),C(z)] -[B(y),C(z)]A(x))F(x,y,z)\\
&\phantom{(A(x)[B(y),}-\left(
B(y)[A(x),C(z)]-[A(x),C(z)]B(y)\right)
F(x,y,z)\\
&=\sum_{i=0}^\infty (-1)^i\binom{r}{i}
\left(
A(x)[B(y),C(z)] - [B(y),C(z)]A(x)\right)
(x-z)^{p+r-i}(y-z)^{q+i}\\
&- \sum_{i=0}^\infty
(-1)^{r+i}\binom{r}{i}\left( B(y)[A(x),C(z)] -
[A(x),C(z)]B(y)\right) (y-z)^{q+r-i}(x-z)^{p+i},
\end{split}
\]
as desired.
\end{proof}
\end{theorem}
Taking $\Resy\Resx$, we have\footnote{This result is
implicitly shown in the work of Li. In fact,
by [Li, proof of Proposition 3.2.9],
we can apply [Li, proof of Proposition
2.2.4] to $a=A(z),b= B(z),c= C(z)$, and
$T = \partial_z$; the result follows(cf. Proposition \ref{theorem:2.9} and
Note \ref{note:5.2}).}
\begin{corollary}[Borcherds
identity for local
fields]\label{corollary:n1} Let
$A(z),B(z)$ and
$C(z)$ be fields on a vector space. 
If they are local, then for
any $p,q,r\in\Z$,
\begin{multline*}
\sum_{i=0}^\infty \binom{p}{i}\left(
A(z)_{(r+i)}B(z)\right)_{(p+q-i)}C(z)\\
= \sum_{i=0}^\infty(-1)^i\binom{r}{i}
\left(
A(z)_{(p+r-i)}(B(z)_{(q+i)}C(z))
-(-1)^rB(z)_{(q+r-i)}(A(z)_{(p+i)}C(z))\right).
\end{multline*}
\end{corollary}

The following special case of the Borcherds
identity holds under a weaker assumption ([Li,
proof of Proposition 3.2.9]).
\begin{proposition}[Li]\label{theorem:2.9}
Let $A(z), B(z)$ and $C(z)$ be fields. If $A(z)$ and $B(z)$ are
local at order $n_0$, then
\[
\sum_{i=0}^\infty
(-1)^i\binom{n}{i}A(z)_{(p+n-i)}(B(z)_{(q+i)}C(z))
 = \sum_{i=0}^\infty
(-1)^{n+i}\binom{n}{i}B(z)_{(q+n-i)}(A(z)_{(p+i)}C(z))
\]
for all $n\geq n_0$.
\end{proposition}
\begin{proof}
By Lemma \ref{lemma:2.1} and \ref{lemma:2.2} 
for $r = n\geq 0$, we have
\[
\begin{split}
&\sum_{i=0}^\infty
(-1)^i\binom{n}{i}[A(x),[B(y),C(z)]](x-z)^{p+n-i}(y-z)^{q+i}\\
&=[A(x),[B(y),C(z)]]F_1(x,y,z)
=[A(x),[B(y),C(z)]]F(x,y,z),\\
&\sum_{i=0}^\infty
(-1)^{n+i}\binom{n}{i}[B(y),[A(x),C(z)]](y-z)^{q+n-i}(x-z)^{p+i}\\
&=[B(y),[A(x),C(z)]]F_2(x,y,z)=[B(y),[A(x),C(z)]]F(x,y,z).
\end{split}
\]
They coincide by the locality of $A(z)$ and $B(z)$.
Take $\Resy\Resz$ to get the result.
\end{proof}

\subsection{Skew symmetry}\label{subsection:2.5}
Let us discuss the identity satisfied by two
fields with respect to the residual products.
We first consider the non-local case:
\begin{proposition}\label{eqn:2.10}
Let $A(z)$ and $B(z)$ be fields. 
Then for any integer $m$, we have
\[
\bigl(B(z)_{(m)}A(z)\bigr)_+
= \sum_{i=0}^\infty
(-1)^{m+i+1}\partial^{(i)}\bigl(
A(z)_{(m+i)}B(z)\bigr)_+.
\]
\end{proposition}
\begin{proof}
Let $n$ be a nonnegative integer. Then
\begin{align*}
&z^n[B(y),A(z)](y-z)^m=
(-1)^{m+1}z^n[A(z),B(y)](z-y)^m\\
&=(-1)^{m+1}\sum_{i=0}^n\binom{n}{i}y^{n-i}
(z-y)^i[A(z),B(y)](z-y)^m\\ &=
(-1)^{m+1}\sum_{i=0}^n(\partial^{(i)}_y
y^n)[A(z),B(y)](z-y)^{m+i}\\
&=(-1)^{m+1}\sum_{i=0}^n
(-1)^{i}y^n\partial_y^{(i)}
\biggl([A(z),B(y)](z-y)^{m+i}\biggr)
+\quad\text{(total derivative)}.
\end{align*}
Take $\Resz\Resy$ of the both sides, to get
\[
\bigl(B_{(m)}A\bigr)_{n}
= \sum_{i=0}^n (-1)^{m+i+1}\partial^{(i)}\bigl(
A_{(m+i)}B\bigr)_{n}.
\]
Since this holds for all nonnegative 
integers $n$, we have the result.
\end{proof}

If the two fields are local, 
then the equality holds also for
the negative parts:

\begin{proposition}\label{proposition:2.11}
Let $A(z)$ and $B(z)$ be fields. 
If $A(z)$ and $B(z)$ are local, then for any
integer $m$, we have,
\[
B(z)_{(m)}A(z)
= \sum_{i=0}^\infty (-1)^{m+i+1}\partial^{(i)}
\bigl(A(z)_{(m+i)}B(z)\bigr).
\]
\end{proposition}

This is a special case of Corollary \ref{corollary:n1}. In fact,
substituting the identity fields $I(z)$ for $C(z)$ and setting $p =
-1, q=0$ and $r = m$ in the corollary and using
Proposition \ref{proposition:1}, 
we have the result (see (\ref{eqn:16}) in
Subsection 3.1). We refer the reader to [K,
p.43] for an alternative proof using the
operator product expansion.

\section{Axioms for a vertex
algebra}\label{section:3} In this section, we
 first review the definition of a vertex
algebra following Borcherds \cite{B92}. We next
consider the inductive structure of the
Borcherds identity, and finally we
describe Li's theorem on a local system of
vertex operators as a direct consequence of
the results of the preceding section.

\subsection{Borcherds' axioms for a vertex
algebra} According to Borcherds \cite{B92}, a
vertex algebra is defined as
follows\footnote{Originally in [B1, Section 4],
the properties (B0),
(\ref{eqn:13}),(\ref{eqn:12}),(\ref{eqn:16})
and (\ref{eqn:19}) given below are taken as the
axioms for a vertex algebra.}:
\begin{definition}\label{definition:6}
A \textit{vertex algebra} is a
vector space $V$ equipped with countably many bilinear
binary operations
\[
\begin{array}{ccl}
V \times V&\longrightarrow &V\\
(a,b)&\longmapsto&a_{(n)}b,\quad (n\in \Z),
\end{array}
\]
and a vector  $\vac \in V$ subject to the
following conditions
\newline\noindent
(B0) For each pair of vectors $a,b\in V$, 
there exists a nonnegative integer $n_0$ such
that	
\[
a_{(n)}b = 0\quad\mbox{for all}\quad n\geq n_0.
\]
\newline
\noindent
(B1) \textbf{(Borcherds identity)}\footnote{This 
identity is nothing but the Cauchy-Jacobi
identity of Frenkel-Lepowsky-Meurman
[FLM, (8.8.29) and (8.8.41)], while the special case
(\ref{eqn:19})  is due to Borcherds \cite{B86}; we here
follow the terminology in \cite{K96}.} 
For all vectors
$a,b,c\in V$ and all integers $p,q,r\in\Z$,
\[
\sum_{i=0}^\infty
\binom{p}{i}(a_{(r+i)}b)_{(p+q -i)}c 
=\sum_{i=0}^\infty (-1)^i\binom{r}{i}\left(
a_{(p+r-i)}(b_{(q + i)}c) -
(-1)^rb_{(q+r-i)}(a_{(p+i)}c)\right).
\]
\newline
\noindent
(B2) For any $a\in V$,
\[
a_{(n)}\vac = \begin{cases}
0,&\quad ( n\geq 0),\\
a,&\quad ( n = -1).
\end{cases}
\]
\end{definition}
The vector $\vac$ is called the vacuum vector of $V$. Note that,
because of (B0), each side of the Borcherds
identity in (B1) is a finite sum. We also note
that (B2) does not contain a condition on
$a_{(n)}\vac$ for $n\leq -2$. However, if
$a_{(-2)}\vac$ are specified for all $a\in V$,
in other words, if we know the endomorphism
\[
\begin{array}{ccc}
T:&V\longrightarrow&V\\
&a\longmapsto&a_{(-2)}\vac,
\end{array}
\]
then $a_{(n)}\vac$ are uniquely determined by $T$ 
as we will see in (\ref{eqn:12}) below.  

It is sometimes convenient to introduce the
generating series
\[
Y(a,z) = \sum_{n\in\Z}a_{(n)}z^{-n-1}
\]
and consider the map
\[
\begin{array}{ccc}
Y:&V\longrightarrow&(\End V)[[z,z^{-1}]]\\
&a\longmapsto&Y(a,z)
\end{array}.
\]
Then the vertex algebra structure is
specified by the triple $(V,\vac, Y)$. 

For the convenience of the reader, we derive
various properties of vertex algebras from the
axioms. First, putting
$a = b = c =\vac$ and $p=q=r=-1$ in (B1), we
immediately see
\begin{equation}\label{eqn:11}
\vac_{(-2)}\vac = 0,\quad \text{i.e.}, \quad T\vac = 0.
\end{equation}
Then the substitution $b = c = \vac$ and $p =0, q= -2, r = n$
shows by (B2)
\[
(a_{(n)}\vac)_{(-2)}\vac = -na_{(n-1)}\vac.
\]
Therefore, for $n\leq -1$, we inductively deduce
\[
a_{(n)}\vac = T^{(-n-1)}a
\]
where $T^{(k)} = T^k/k!$. 
Thus (B2) is completed as
\begin{equation}\label{eqn:12}
a_{(n)}\vac = 
\begin{cases}
0,&\quad (n\geq 0),\\
T^{(-n-1)}a,&\quad (n \leq -1).
\end{cases}
\end{equation}
Now, let $b = c = \vac, p=-1,q=n, r=0$ in (B1). Then,
by (B2) and (\ref{eqn:11}), we have
\begin{equation}\label{eqn:13}
\vac_{(n)}a = 
\begin{cases}
0,&\quad (n\neq -1),\\
a,&\quad (n = -1).
\end{cases}
\end{equation}
Next, let $b = \vac, p=n,q=0,r=-2$ in (B1) and replace
$c$ by
$b$. Then by (\ref{eqn:13}),
\begin{equation}\label{eqn:14}
(Ta)_{(n)}b = -na_{(n-1)}b.
\end{equation}
Further, $c = \vac,p = 0, q= -2, r=n$ yields by (B2)
\begin{equation}\label{eqn:15}
a_{(n)}(Tb) = T(a_{(n)}b) + na_{(n-1)}b.
\end{equation}
Namely,
$[T,a_{(n)}] = -na_{(n-1)}$, called the
translation covariance. Then comparing
(\ref{eqn:14}) and (\ref{eqn:15}) we see that
$T$ is a derivation for all the products:
\[
T(a_{(n)}b) = (Ta)_{(n)}b + a_{(n)}(Tb).
\]
Finally, $c = \vac, p = -1, q= 0, r =n$ gives rise to
\begin{equation}\label{eqn:16}
b_{(n)}a  =  
\sum_{i=0}^\infty (-1)^{n+i+1}T^{(i)}(a_{(n+i)}b)
\end{equation}
which is called the \textit{skew symmetry}\,.

In terms of the generating series $Y(a,z)$,
the properties (\ref{eqn:12})--(\ref{eqn:16})
are rewritten as follows:
\begin{align*}
&Y(a,z)\vac = e^{Tz}a,\quad
Y(\vac,z)  = \vac,\quad
Y(Ta,z)= \partial_z Y(a,z),\\
&[T,Y(a,z)]= \partial_z Y(a,z),\quad
Y(b,z)a= e^{Tz}Y(a,-z)b.
\end{align*}
\begin{remark}\label{remark:3.1.2}
Let $V$ be a vertex algebra satisfying the
following condition: For any $a\in V$, there
exists a nonnegative integer $n_0$ such that
$a_{(n)}V = 0$ for all $n\geq n_0$. Then, by
(\ref{eqn:15}) we actually have $a_{(n)}V = 0$
for all $n\geq 0$.
\end{remark}
\begin{note}\label{note:3.1.3}
The structure of a vertex algebra as in the
remark is described as follows (\cite{B86}):
It has a structure of commutative algebra with
respect to the multiplication defined by $ab =
a_{(-1)}b$, and $T$ is a derivation.
Conversely, any commutative algebra with a
derivation $T$ has a vertex algebra structure
with respect to
\[
a_{(n)}b = 
\begin{cases}
0&\quad (n\geq 0),\\
(T^{(-n-1)}a)b&\quad (n\leq -1).
\end{cases}
\]
Any finite-dimensional vertex algebra is
described in this way ([B2, p.416]), since it
obviously satisfies the condition in the remark.
\end{note}

\subsection{Structure of Borcherds identity}
Let $B(p,q,r)$ denote one of the three terms of
the Borcherds identity:
\begin{align*}
B(p,q,r)&=
 \sum_{i=0}^\infty
\binom{p}{i}(a_{(r+i)}b)_{(p+q -i)}c, \quad
\sum_{i=0}^\infty
(-1)^i\binom{r}{i} a_{(p+r-i)}(b_{(q +
i)}c),\\
&\quad\quad\text{or}\quad\sum_{i=0}^\infty
(-1)^{r+i}\binom{r}{i}b_{(q+r-i)}(a_{(p+i)}c).
\end{align*}
Then, straightforward calculation shows 
\begin{equation}\label{eqn:17}
B(p+1,q,r) = B(p,q+1, r) + B(p,q,r+1).
\end{equation}
Therefore, the Borcherds identity for two of the
indices $(p+1,q,r), (p,q+1,r)$ and $(p,q,r+1)$
imply the Borcherds identity for the other index.
This proves

\begin{proposition}\label{proposition:3}
The Borcherds identity for all $p$ and $q$ with
fixed
$r$ and for all
$q$ and $r$ with fixed $p$ imply the Borcherds
identity for all $p,q$ and $r$.
\end{proposition}

Now let us consider special cases of
the Borcherds identity. Setting $r = 0$ and $p =
0$ respectively, we have
\begin{equation}\label{eqn:18}
[a_{(p)},b_{(q)}] 
= \sum_{i=0}^\infty
\binom{p}{i}(a_{(i)}b)_{(p+q-i)}
\end{equation}
and 
\begin{equation}\label{eqn:19}
(a_{(r)}b)_{(q)}  = \sum_{i=0}^\infty
(-1)^i\binom{r}{i}(a_{(r-i)}b_{(q+i)} - (-1)^r
b_{(q+r-i)}a_{(i)})
\end{equation}
called the \textit{commutator formula} and the
\textit{associativity formula}.
Here we have omitted the overall $c$. We next
take an
$r_0$ such that $a_{(r)}b = 0$ for all $r\geq
r_0$. Then for such $r$, we have,
\begin{equation}\label{eqn:20}
\sum_{i=0}^\infty
(-1)^i\binom{r}{i}\left( a_{(p+r-i)}b_{(q+i)} - (-1)^rb_{(q+r
-i)}a_{(p+i)}\right) = 0,\quad (r\geq r_0)
\end{equation}
which is nothing else but the \textit{locality} or
the \textit{commutativity}.
Finally, take a $p_0$ such that $a_{(p)}c = 0$ for all $p\geq
p_0$. Then, for such $p$, we have
\begin{equation}\label{eqn:21}
\sum_{i=0}^\infty\binom{p}{i}(a_{(r+i)}b)_{(p+r-i)}c=
\sum_{i=0}^\infty(-1)^i\binom{r}{i}
a_{(p+r-i)}(b_{(q+i)}c),\quad (p\geq p_0)
\end{equation}
called the \textit{duality} or
the \textit{associativity}. This time we can not
omit
$c$ because the choice of $p_0$ depends on it.

The relations (\ref{eqn:18})--(\ref{eqn:21}) are
rewritten in terms of generating series as:
\begin{align*}
&[a_{(p)},Y(b,z)] =\sum_{i=0}^\infty
\binom{p}{i}Y(a_{(i)}b,z)z^{p-i},\\
&Y(a_{(r)}b,z) = Y(a,y)_{(r)}Y(b,z),\\
&Y(a,y)Y(b,z)(y-z)^r =
Y(b,z)Y(a,y)(y-z)^r,\quad (r\geq r_0),\\
&Y(Y(a,y)b,z)(y+z)^pc= Y(a,y+z)Y(b,z)(y+z)^pc
,\quad (p\geq p_0).
\end{align*}

As an immediate consequence of Proposition
\ref{proposition:3}, we have

\begin{proposition}\label{proposition:3.2}
The axiom (B1) is equivalent to either (\ref{eqn:18}) or
(\ref{eqn:20}) and either (\ref{eqn:19}) or (\ref{eqn:21}).
\end{proposition}
\medskip
We here note the remarkable correspondence
between the coefficients and indices in the
Borcherds identity and the coefficients and
exponents of the expansions
\begin{align*}
(x-&y)^r(y-z)^q(x-z)^p = \sum_{i=0}^\infty
\binom{p}{i}(x-y)^{r+i}(y-z)^{p+q-i},\\
&\sum_{i=0}^\infty(-1)^i
\binom{r}{i}(x-z)^{p+r-i}(y-z)^{q+i},
\quad\text{and}\quad\sum_{i=0}^\infty(-1)^{r+i}
\binom{r}{i}(y-z)^{q+r-i}(x-z)^{p+i}.
\end{align*}
With this resemblance in mind, we see the relation
(\ref{eqn:17}) corresponds to
\[
\begin{split}
(x-y)^r&(y-z)^q(x-z)^{p+1} \\ 
&= (x-y)^r(y-z)^{q+1}(x-z)^{p}
+(x-y)^{r+1}(y-z)^q(x-z)^{p}.
\end{split}
\]

\subsection{Li's theorem}
We now return to the situation of Section 2. 
Let $M$ be a vector
space, and let $A(z), B(z)$ and $C(z)$ be local
fields on $M$, while
$I(z)$ the identity field. 
Then we have already shown that there exists a
nonnegative integer $n_0$ such that
$A(z)_{(n)}B(z) = 0$ for all $n\geq n_0$
(\ref{eqn:1.12}),
that the Borcherds identity holds (Corollary
\ref{corollary:2.1}):
\[
\begin{split}
&\sum_{i=0}^\infty \binom{p}{i}(A(z)_{(r+i)}B(z))_{(p+q -i)}C(z) \\
&=\sum_{i=0}^\infty (-1)^i\binom{r}{i}\left(
A(z)_{(p+r-i)}(B(z)_{(q + i)}C(z)) -
(-1)^rB(z)_{(q+r-i)}(A(z)_{(p+i)}C(z))\right),
\end{split}
\]
and that we have (\ref{eqn:9})
\[
A(z)_{(n)}I(z) = 
\begin{cases}
0&\quad (n\geq 0),\\
A(z)&\quad (n = -1).
\end{cases}
\]
\medskip
They are precisely Borcherds' axioms for a vertex
algebra. Therefore, we have established 
the following theorem [Li, Theorem
3.2.10]\footnote{Li stated the result only for
a maximal pairwise local space of fields.
However, his proof indeed applies to the
statement in the theorem.}.
\begin{theorem}[Li]\label{theorem:3.3}
Let $M$ be a vector space, and let
$\mathcal{O}$ be a vector space consisting of
fields on $M$ which are pairwise local. If
$\mathcal{O}$ is closed under the residual
products and contains the identity field
$I(z)$, then the residual products equip $V
=\mathcal{O}$ with a structure of vertex
algebra with the vacuum vector $\vac = I(z)$.
\end{theorem}

In particular, let $\cals$ be a set of pairwise
local fields and let $\calo$ be the space
$\langle\cals\,\rangle$ generated by $\cals$.
Then $\calo$ is closed under the residual
products and contains the identity field; it is
a vertex algebra. It follows from the
associativity formula (\ref{eqn:19}) and the
formulas  (\ref{eqn:6}), (\ref{eqn:7})
and (\ref{eqn:8}) that the space
$\langle\cals\,\rangle$ is the linear span of
the fields of the form
\[
A^{(\lambda_1)}(z)_{(n_1)}A^{(\lambda_2)}(z)_{(n_2)}
\cdots A^{(\lambda_\ell)}(z)_{(n_\ell)}I(z)
\]
where
$
n_1,\dots, n_{\ell-1}\in\Z, n_{\ell}\in
\Z_{<0}, 
A^{(\lambda_1)}(z), \dots,
A^{(\lambda_\ell)}\in\cals.
$
Here we understand the nested products as 
$A(z)_{(m)}B(z)_{(n)}C(z)
=A(z)_{(m)}(B(z)_{(n)}C(z))$.

\begin{remark}\label{remark:3.3.2}
Let $V_0$ be a vector space and suppose given
a series of binary operations satisfying (B0)
and (B1). Then the image $\calv_{Y_0}$ of the 
corresponding map $Y_0:V_0\longrightarrow 
(\End V_0)[[z,z^{-1}]]$ is pairwise
local by (\ref{eqn:20}). Therefore the space
$V = \langle\calv_{Y_0}\rangle$ generated by
$\calv_{Y_0}$ is a vertex algebra with respect
to the residual products by the theorem. Thus
we have constructed a canonical map from $V_0$
to a vertex algebra $V$, which preserves the
binary operations by (\ref{eqn:19}). In
particular, if $Y_0$ is injective, then $V_0$
is canonically embedded in a vertex algebra.
\end{remark}

\begin{note}\label{note:3.1}
Li's original proof of the above theorem is as
follows.
For each $A(z)\in\mathcal{O}$, consider
\[
Y(A(z),\zeta) =
\sum_{n\in\Z}A(z)_{(n)}\zeta^{-n-1}\in
(\End\mathcal{O})[[\zeta,\zeta^{-1}]].
\]
Then he shows that the map
$
Y:\mathcal{O}\longrightarrow (\End
\mathcal{O})[[\zeta,\zeta^{-1}]]
$
satisfies Goddard's axioms (G0)--(G3) (see
Subsection 5.2) where
$V=
\mathcal{O}, \vac = I(z)$ and
$T = \partial_z$. (In particular, the locality
(G1) follows from one of his result which we have
explained in Proposition \ref{theorem:2.9}.) Then by
the equivalence of Goddard's axioms and
Borcherds' axioms, which he shows by direct
calculation (cf. Note \ref{note:5.2}), the
result follows.
\end{note}

\section{State-field correspondence}
In this section, we will describe the
state-field correspondence of a vertex algebra
and the characterization of vertex algebras due
to Lian-Zuckerman.  Throughout this section, we
suppose given a vector space $V$ and a nonzero
vector
$\vac\in V$.

\subsection{Creative fields}\label{subsection:4.1}
We begin with the notion of creative fields as
follows:

\begin{definition}
A field $A(z)$ on $V$ is \textit{creative} 
with respect to
$\vac$ if $A(z)\vac\in V[[z]]$.
\end{definition}

In other words, $A(z)$ is
creative if and only if $A_n\vac
= 0$ for all $n\geq 0$.

We define $\tilde{\mathcal{O}} = \tilde{\mathcal{O}}(V,\vac)$ to be
the set of all creative fields 
on $V$ with respect to $\vac$.
For a creative field $A(z)\in
\tilde{\mathcal{O}}$, we set 
\[
\ket{A} = \underset{z\rightarrow 0}{\lim}(A(z)\vac) 
= A_{-1}\vac
\] and call it the \textit{state} corresponding to the field $A(z)$.

Consider the map 
\[
\begin{array}{cccc}
s:&\tilde{\mathO}&\longrightarrow&V\\
&A(z)&\longmapsto&\ket{A}
\end{array}
\]
which assigns the state to a creative field.

\begin{lemma}\label{lemma:4.1}
The space $\tilde{\mathO}$ is closed under the residual products and
we have
\[
s(A(z)_{(m)}B(z)) = A_{m}\ket{B}
\]
for all $A(z),B(z)\in\tilde{\mathO}$.
\end{lemma}
\begin{proof}
For $A(z),B(z)\in\tilde{\mathO}$,
we have
\[
\begin{split}
\biggl( A(z)_{(m)}B(z)\biggr)\vac
& = \sum_{n\in\Z}\biggl(
\sum_{i=0}^\infty
(-1)^i\binom{m}{i}\bigl(A_{m-i}B_{n+i}
-(-1)^mB_{m+n-i}A_i\bigr)\biggr)\vac z^{-n-1}\\ 
& =\sum_{n\leq -1}\biggl(
\sum_{i=0}^{-n-1} (-1)^i\binom{m}{i}A_{m-i}B_{n+i}\vac
\biggr) z^{-n-1}.
\end{split}
\]
Hence $A(z)_{(m)}B(z)$ belongs to $\tilde{\mathO}$ and we have
$
s(A(z)_{(m)}B(z)) = \underset{z\rightarrow
0}{\lim} (A(z)_{(m)}B(z)\vac) = A_mB_{-1}\vac =
A_m\ket{B}.
$
\end{proof}

The following lemma, which will be used in the next
section, is a part of the statement known as
Goddard's uniqueness theorem:
\begin{lemma}\label{lemma:4.2}
Let $\calv$ be a subspace of $\tilde{\mathO}$ which
is pairwise local. Then, if the map $s|_{\calv}:\calv
\rightarrow V$ is surjective, then the map
\[
\begin{array}{cccc}
s|_{\calv}:&V&\rightarrow&V[[z]]\\
&A(z)&\mapsto&A(z)\vac
\end{array}
\]
is injective.
\end{lemma}
\begin{proof}
Suppose $A(z)\vac = 0$ and take any $u\in V$. Since 
$\scalv$ is surjective, there is a field $U(z)\in \calv$
such that $\ket{U} = s(U(z)) = u$. 
Then, by the locality,
\[
z^nA(z)u = \limit{y}{0}(z-y)^nA(z)U(y)\vac
= \limit{y}{0}(z-y)^nU(y)A(z)\vac= 0
\]
for sufficiently large $n$. Since $u$ is as
arbitrary, we have $A(z) = 0$.
\end{proof}

\subsection{State-field
correspondence}\label{subsection:4.2}
Let us consider the case when $V$ has a structure of
vertex algebra with the vacuum vector $\vac$. Consider
the generating series
$Y(a,z) = \sum_{n\in\Z}a_{(n)}z^{-n-1}$
and put 
\[
\calv_{Y} = \{Y(a,z)\,|\,a\in V\}.
\]
Then the axiom (B2) says that $\calv_Y$ is contained in
$\tilde{\calo}$ and the map 
\[
\begin{array}{cccc}
Y:&V&\longrightarrow&\calv_Y\\
&a&\longmapsto&Y(a,z)
\end{array}
\]
is the inverse of
\[
\rest{s}{\calv_Y}:\calv_Y\longrightarrow V.
\]
In particular, they are isomorphisms of vector spaces.

\begin{remark}\label{remark:4.3}
Since the map $Y$ is recovered from the space $\calv_Y$
as the inverse of $\rest{s}{\calv_Y}$, a vertex algebra
structure $(V,\vac,Y)$ is uniquely determined by the
subspace $\calv_Y\subset \tilde{\calo}$
\end{remark}

Now, we have
$Y(a_{(n)}b,z) = Y(a,z)_{(n)}Y(b,z)$
by (\ref{eqn:19}) and
$Y(\vac,z) = I(z)$ by (\ref{eqn:13}).
Therefore, we obtain the following theorem:
\begin{theorem}[State-field
correspondence]\label{theorem:4.4}
Let $(V,\vac,Y)$ be a vertex algebra. Then the
residual products equip $\calv_Y$ with a structure
of vertex algebra with the vacuum vector  being  the
identity field $I(z)$ such that the map
\[
\rest{s}{\calv_Y}:\calv_Y\longrightarrow V
\]
is an isomorphism of vertex algebras.
\end{theorem}

Note that the formula (\ref{eqn:14}) means that
the differentiation $\partial_z$ corresponds to
the translation operator $T$ under the
state-field correspondence.

\subsection{Characterization of the image
 (I)}\label{subsection:4.3}

Let us characterize vertex algebras on $V$
with the vacuum vector $\vac$ by means of the subspaces
$\calv_Y =\{Y(a,z)|a\in V\}\subset \tilde{\calo}$.

We consider the following set of conditions on a
subspace 
$\calv\subset\tilde{\calo}$:
\begin{description}
\item[(L1)]
$\mathcal{V}$ is pairwise local.
\item[(L2)]
$\mathcal{V}$ is closed under the residual products.
\item[(L3)]
The map
$s|_{\mathcal{V}}:\mathcal{V}\rightarrow
V$ is an isomorphism of vector spaces.
\end{description}

\begin{proposition}\label{proposition:4.4}
If a subspace $\calv\subset \tilde{\calo}$
satisfies (L1)--(L3), then the residual products
equip $\calv$ with a structure of vertex algebra
with the vacuum vector being the identity field
$I(z)$.
\end{proposition}
\begin{proof}
Let $\calv\subset\tilde{\calo}$ satisfy
(L1)--(L3). Then, since $\calv$ is pairwise local
by (L1), the residual products satisfy (B0) and
(B1) by (\ref{eqn:1.12}) and Corollary
\ref{corollary:n1}. Now let $I(z)$ be the unique
field in $\calv$ such that $s(I(z)) = \vac$. Then,
for any $A(z)\in\calv$, we have 
\[
s(A(z)_{(n)}I(z)) = A_n\vac = 
\begin{cases}
0&\quad (n\geq 0),\\
\ket{A}&\quad(n=-1)
\end{cases}
\]
by Lemma \ref{lemma:4.1} and the creativity of
$A(z)\in
\calv$. Since $A(z)_{(n)}I(z)\in\calv$ by (L2), it
follows from (L3) that
\[
A(z)_{(n)}I(z) =
\begin{cases}
0&\quad (n\geq 0),\\
A(z)&\quad(n=-1),
\end{cases}
\]
which is nothing else but (B2). Hence the residual
products equip $\calv$ with the structure of vertex
algebra with the vacuum vector being $I(z)$. In
particular, by (\ref{eqn:13}),
\[
I_n\ket{A} = s(I(z)_{(n)}A(z)) = 
\begin{cases}
0&\quad (n\neq -1),\\
\ket{A}&\quad(n=-1).
\end{cases}
\]
Since $\ket{A}$ is arbitrary in $V$, we see that
$I(z)$ is the identity field.
\end{proof}

Therefore, since we have the isomorphism
\[
\rest{s}{\calv}:\calv\longrightarrow V
\]
by (L3), we can introduce a vertex algebra
structure on $V$ through the isomorphism:
\[
\ket{A}_{(n)}\ket{B} = s(A(z)_{(n)}B(z)).
\]
Then by Lemma \ref{lemma:4.1} we have
\[
\ket{A}_{(n)}\ket{B} =
A_n\ket{B},\quad\text{i.e.,}\quad Y(\ket{A},z) =
A(z).
\]
Hence the map
$Y$ of the vertex algebra structure on
$V$ coincides with the inverse of
$\rest{s}{\calv}:\calv\longrightarrow V$ and we have
$\calv = \calv_Y$.

Thus we have established the following
theorem stated by
Lian-Zuckerman\footnote{To be precise,
Lian-Zuckerman assumed in addition that 
$\calv$ contains the identity field. However,
it follows from the other
assumptions as we saw in Proposition
\ref{proposition:4.4}} [LZ2, Theorem 5.6]:

\begin{theorem}\label{theorem:4.5}
Let $V$ be a vector space, and let $\vac\in V$
be a nonzero vector. If a subspace
$\calv\subset\tilde{\calo}$ satisfies (L1)--(L3),
then there exists a unique vertex algebra structure
on
$V$ with the vacuum vector $\vac$ such that $\calv =
\calv_Y$, where the map $Y$ is given by the inverse
of $\rest{s}{\calv}$.
\end{theorem}

Conversely, if $(V,\vac,Y)$ is a vertex algebra,
then the space $\calv= \calv_Y\subset\tilde{\calo}$
satisfies (L1)--(L3), Therefore, the conditions
(L1)--(L3) on a subspace
$\calv\subset\tilde{\calo}$ characterize vertex
algebra structures on $V$ with the vacuum
vector $\vac$. More precisely,
\begin{corollary}\label{corollary:4.6}
The correspondence 
$Y\longmapsto \calv_Y = \{\,Y(a,z)\,|a\in
V\,\}$ from the set of vertex algebra structures
on $V$ with the vacuum vector $\vac$ to the set
of the subspaces
$\calv\subset\tilde{\calo}$ satisfying (L1)--(L3)
is bijective, and the inverse correspondence is
given by
$\calv\longmapsto Y_\calv =
(\rest{s}{\calv})^{-1}$.
\end{corollary}

In  other words, giving a vertex algebra structure
$(V,\vac,Y)$ is equivalent to giving a subspace
$\calv\subset\tilde{\calo}$ satisfying (L1)--(L3).

\section{Goddard's axioms and existence
theorem}
In this section, we will describe the
characterization of vertex algebras by the
axioms essentially given by Goddard \cite{G},
where the translation operator
$T$ plays an essential role. We also describe the
existence theorem due to
Frenkel-Kac-Radul-Wang as a consequence of
still another characterization.
Throughout this section again, we suppose given a
vector space $V$ and a nonzero vector
$\vac\in V$.

\subsection{Translation covariance}
Let us first consider the role played by the
translation operator. Suppose given an endomorphism
$T:V\longrightarrow V$ such that $T\vac = 0$.

\begin{definition}\label{definition:5.1}
A field $A(z)$ is \textit{translation covariant} with
respect to $T$, if
\begin{equation}\label{eqn:24}
[T,A(z)] = \partial A(z)
\end{equation}
is satisfied.
\end{definition}
\begin{remark}\label{remark:5.1}
The property (\ref{eqn:24}) is equivalent to
\[
e^{yT}A(z)e^{-yT} = \rest{A(y+z)}{|y|<|z|},
\]
where the right-hand side means 
\[
\rest{A(y+z)}{|y|<|z|} =
\sum_{n\in\Z}A_n\rest{(y+z)^n}{|y|<|z|} =
\sum_{n\in\Z}\sum_{i=0}^\infty\binom{n}{i}A_ny^iz^{n-i}.
\]
\end{remark}

We say that a set of fields $\cals$ is translation
covariant if all the fields in $\cals$ are
translation covariant with respect to the
same $T$.

\begin{lemma}\label{lemma:5.1}
If two fields $A(z)$ and $B(z)$ are translation
covariant, then so are the residual products
$A(z)_{(n)}B(z)$.
\end{lemma}
\begin{proof}
It is obvious by 
$[T,A(z)_{(n)}B(z)] = [T,A(z)]_{(n)}B(z)
+A(z)_{(n)}[T,B(z)]$ \linebreak and
$\partial(A(z)_{(n)}B(z)) = \partial
A(z)_{(n)}B(z)+ A(z)_{(n)}\partial B(z)$.
\end{proof}

Now consider the map
\[
\begin{array}{cccc}
\imath:&V[[z]]&\longrightarrow&V\\
&u(z)&\longmapsto&u(0)
\end{array}
\]
which assigns the initial value $u(0) = u_{-1}$ to a
series $u(z) = \sum_{n\leq -1}u_n z^{-n-1}$. 
If a field $A(z)$ is creative and translation
covariant, then the series $A(z)\vac\in V[[z]]$
satisfies the differential equation
\begin{equation}\label{eqn:25}
\partial_z(A(z)\vac) = T(A(z)\vac)
\end{equation}
whose solution is uniquely determined by the initial
value $\imath(A(z)\vac) = \ket{A}$ (cf. [K, Remark
4.4]).
More precisely,
\begin{lemma}\label{lemma:5.2}
If a space of fields $\calv$ is creative and
translation covariant, then the map
\[
\begin{array}{cccc}
\rest{\imath}{\calv\vac}:&\calv\vac&\longrightarrow&V\\
&A(z)\vac&\longmapsto&\ket{A}
\end{array}
\]
is injective.
\end{lemma}
\begin{proof}
By the assumptions, we have 
\[
\sum_{i=0}^\infty iA_{-i-1}\vac z^{i-1} =
\partial A(z)\vac=[T,A(z)]\vac
=TA(z)\vac= \sum_{i=0}^\infty TA_{-i-1}\vac
z^{-i}.
\]
Equating the coefficients,
$
(i+1)A_{-i-2}\vac = TA_{-i-1}\vac, (i\geq 0)
$.
Therefore, if $\ket{A} = 0$, then we inductively
deduce
$A_{-i-1}\vac = 0, (i = 0,1,\dots)$,
namely $A(z)\vac = 0$.
\end{proof}
\begin{remark}\label{remark:5.2}
If $A(z)$ is creative and translation covariant,
then, by solving the differential equation
(\ref{eqn:25}), we have 
$
A(z)\vac = e^{zT}\ket{A}
$.
\end{remark}

\subsection{Goddard's axioms}\label{subsection:5.2}
Suppose given a linear map
\[
\begin{array}{cccc}
Y:&V\longrightarrow&(\End V)[[z,z^{-1}]]\\
&a\longmapsto&Y(a,z)
\end{array},
\]
and consider the following set of axioms\footnote{
These are essentially the axioms considered by
Goddard
\cite{G}. More precisely, he assumed 
$Y(a,z)\vac = e^{zT}a$ instead of $[T,Y(a,z)] =
\partial Y(a,z)$ in (G3). However, they are
equivalent under the axioms (G0)--(G2) by
Lemma \ref{lemma:4.2} and Remark
\ref{remark:5.2}. (cf. [FKRW, Section 3], [K,
Subsection 1.3].)} on
$(V,\vac,Y)$:
\begin{description}
\item[(G0)]
For any $a,b\in V$, $Y(a,z)b\in V((z))$.
\item[(G1)]
For any $a,b\in V$, $Y(a,z)$ and $Y(b,z)$ are local.
\item[(G2)]
For any $a\in V$, $Y(a,z)\vac$ is a formal power
series with the constant term $a$.
\item[(G3)]
There exists an endomorphism $T:V\longrightarrow V$ such
that $T\vac = 0$ and $[T,Y(a,z)] = \partial Y(a,z)$ for
all $a\in V$.
\end{description}

Let $(V,\vac,Y)$ satisfy the axioms (G0)--(G3). For
each $a\in V$, we define the endomorphism
$a_{(n)}\in\End V$ by the expansion
$
Y(a,z) = \sum_{n\in\Z}a_{(n)}z^{-n-1}
$.
Now, let $a,b\in V$, and consider the series
$Y(a,z)_{(n)}Y(b,z)$ and $Y(a_{(n)}b,z)$. Then
they are creative and translation covariant.
Since
\[
\limit{z}{0}Y(a,z)_{(n)}Y(b,z)\vac = a_{(n)}b =
\limit{z}{0}Y(a_{(n)}b,z)\vac,
\]
we have $Y(a,z)_{(n)}Y(b,z)\vac=Y(a_{(n)}b,z)\vac$
by lemma \ref{lemma:5.2}. Then by Lemma
\ref{lemma:4.2} applied to $\calv =
\{\,Y(a,z)\,|\,a\in V\,\}$, we obtain the
associativity formula
\[
Y(a,z)_{(n)}Y(b,z)=Y(a_{(n)}b,z),
\]
which, together with the locality, implies
the Borcherds identity (B1) 
\[
\sum_{i=0}^\infty \binom{p}{i}\left(
a_{(r+i)}b\right)_{(p+q-i)}c
= \sum_{i=0}^\infty
(-1)^i\binom{r}{i}
\left(
a_{(p+r-i)}(b_{(q+i)}c)
- (-1)^rb_{(q+r-i)}(a_{(p+i)}c)
\right).
\]
by Proposition
\ref{proposition:3.2}.

The other implications are easy, and we have (cf.
 [Li, Proposition 2.2.4], [K,
Proposition 4.8]):

\begin{theorem}
\label{theorem:5.4}
Let $V$ be a vector space, and let $\vac \in V$
be a nonzero vector. Then, a linear map
\[
Y:V\longrightarrow (\End V)[[z,z^{-1}]]
\]
gives rise to a vertex algebra structure on $V$ with
the vacuum vector $\vac$ if and only if $(V,\vac,Y)$ 
satisfies Goddard's axioms (G0)--(G3).
\end{theorem}
\begin{note}\label{note:5.2}
Li's proof of this result is as follows. Let
$(V,\vac,T)$ satisfy (G0)--(G3). First note that
$
e^{zT}Y(a,y) = Y(a,y+z)e^{zT}
$
by the translation covariance (G3). Then, for any
$a,b\in V$, we have
\[
\begin{split}
(y-z)^m&Y(b,z)Y(a,y)\vac=(y-z)^mY(a,y)Y(b,z)\vac\\
&=(y-z)^mY(a,y)e^{zT}b=(y-z)^me^{zT}Y(a,y-z)b
\end{split}
\]
for $m\gg 0$. Letting $y=0$ and 
dividing the both sides by
$(-z)^m$, we have
$
Y(b,z)a = e^{zT}Y(a,-z)b
$.
Then for any $a,b\in V$, we have 
\[
\begin{split}
&(x+z)^mY(a,x+z)Y(b,z)c  \\
&=(x+z)^mY(a,x+z)e^{zT}Y(c,-z)b
=(x+z)^me^{zT}Y(a,x)Y(c,-z)b\\
&=(x+z)^me^{zT}Y(c,-z)Y(a,x)b
=(x+z)^mY(Y(a,x)b,z)c
\end{split}
\]
for $m\gg 0$, which is the duality (cf. [G, Theorem 3]).
Finally, Li shows by calculation involving the delta
function that the locality and the duality imply
the Borcherds identity.
This last step is simplified by our consideration in
Subsection 3.2 (cf. Proposition
\ref{proposition:3.2}).
\end{note}

\subsection{Characterization of the image 
(II)}\label{subsection:5.3}
In Subsection 4.3, we gave a set of conditions on
$\calv\subset\tilde{O}$ to be the image $\calv_Y$ of
a vertex algebra $(V,\vac,Y)$. Let us rewrite them
using the translation covariance.

To this end, let us consider the map
$\rest{s}{\calv}:\calv\longrightarrow V$.
It decomposes as
\[
\begin{array}{ccccc}
\calv&\longrightarrow&\calv\vac&\longrightarrow&V\\
A(z)&\longmapsto&A(z)\vac&\longmapsto&\ket{A}.
\end{array}
\]
Then by combining Lemma \ref{lemma:4.2} and Lemma
\ref{lemma:5.2}, we have
\begin{lemma}\label{lemma:5.5}
Let $\calv$ be a subspace of $\tilde{\calo}$. If the
map $\rest{s}{\calv}:\calv\longrightarrow V$ is
surjective, and if $\calv$ is pairwise local and
translation covariant, then the map
$\rest{s}{\calv}$ is bijective.
\end{lemma}

The following proposition illustrates the role of
$T$.

\begin{proposition}\label{proposition:5.6}
If $\calv\subset \tilde{\calo}$ is pairwise local
and the map $\rest{s}{\calv}:\calv\longrightarrow V$
is surjective, then the following conditions are
equivalent:
\begin{description}
\item[(a)]
$\calv$ is closed under the residual
products and the map
$\rest{s}{\calv}:\calv\longrightarrow
V$ is bijective.
\item[(b)]
$\calv$ is translation covariant.
\end{description}
\end{proposition}
\begin{proof}
Suppose that the condition (a) holds. Then $\calv$
satisfies (L1)--(L3), and it coincides with the
image $\calv_Y$ of a vertex algebra $(V,\vac,Y)$ by
Theorem \ref{theorem:4.5}. Hence $\calv$ is
translation covariant.

Conversely, suppose that (b) holds. Then the map
$\rest{s}{\calv}:\calv\longrightarrow V$ is
bijective by Lemma \ref{lemma:5.5}. Now, let
$\tilde{\calv}= \langle\calv\rangle$ be the
space of fields generated by $\calv$. Then,
$\tilde{\calv}$ is pairwise local and
translation covariant, so we have
$\rest{s}{\tilde{\calv}}:\tilde{\calv}\longrightarrow
V$ is also bijective. Therefore $\calv =
\tilde{\calv}$ and $\calv$ is closed under the
residual products.
\end{proof}

Thus we are led to the following set of
conditions on a subspace $\calv\subset \tilde{O}$:
\begin{description}
\item[(T1)] $\calv$ is pairwise local.
\item[(T2)] The map $\rest{s}{\calv}:\calv\longrightarrow
V$ is surjective.
\item[(T3)] $\calv$  is translation covariant.
\end{description}

Then the above proposition says that the (T1)--(T3)
is equivalent to the (L1)--(L3). Therefore, by
Theorem \ref{theorem:4.5},
\begin{theorem}\label{theorem:5.7}
Let $V$ be a vector space, and let $\vac\in V$
be a nonzero vector. If a subspace
$\calv\subset\tilde{\calo}$ satisfies (T1)--(T3),
then there exists a unique vertex algebra structure
on
$V$ with the vacuum vector $\vac$ such that $\calv =
\calv_Y$, where the map $Y$ is given by the inverse of
$\rest{s}{\calv}:\calv\longrightarrow V$.
\end{theorem}

\subsection{Existence theorem}\label{subsection:4.6}
An immediate consequence of the last
theorem is the following existence theorem
([FKRW, Proposition 3.1], [K, Theorem 4.5]):
\begin{theorem}[Existence theorem]\label{theorem:}
Let $V$ be a vector space, and let $\vac\in V$
be a nonzero vector. If a subset $\cals\subset
\tilde{\calo}$ satisfies 
\begin{description}
\item[(S1)]
$\cals$ is pairwise local.
\item[(S2)]
The set
$\{A^{\lambda_1}_{-j_1-1}
\cdots A^{\lambda_k}_{-j_k-1}\vac\,|
k,j_1,\dots,j_k\in\N, A^{\lambda_1}(z),\dots,
A^{\lambda_k}(z)\in \cals\,\}$
spans $V$.
\item[(S3)]
$\cals$ is translation covariant for some
endomorphism
$T:V\longrightarrow V$.
\end{description}
Then there exists a unique vertex algebra structure
on $V$ with the vacuum vector $\vac$ such that
$Y(\ket{A^\lambda},z) = A^\lambda(z)$
for any $A^\lambda(z)\in\cals$.
\end{theorem}
Just apply the theorem to $\calv=
\langle\cals\,\rangle$ to get the corollary. We
note that the map $Y$ is given by 
\[
Y(A^{\lambda_1}_{-j_1-1}
\cdots A^{\lambda_k}_{-j_k-1}\vac,z)
= \NO \partial^{(j_1)}A^{\lambda_1}(z)
\cdots \partial^{(j_k)}A^{\lambda_k}(z)\NO,
\]
which span the space $\calv$.

\section{Analytic method}\label{section:6}
In this section, we discuss
the condition sufficient for justifying the
argument of contour deformation in
two-dimensional quantum field theory. We
define the notion of admissible fields, and
apply it to the Borcherds identity for local
fields and that of vertex algebra respectively.

In this section, we will work over the complex number
field $\C$, while some of the statements make sense
over any field of characteristic zero.

\subsection{Admissible fields}
Let $M$ be a $\C$-vector space and $M^*$ the
dual space of $M$. We denote the canonical
pairing by
\[
\langle\,,\,\rangle:M^*\times M\longrightarrow \C.
\]

We say that a subspace $M^\vee\subset M^*$ is
\textit{nondegenerate} if the
condition $\pairing{M^\vee}{u} = 0$ implies $u=0$.

\begin{lemma}\label{lemma:6.1.1}
Let $N_m$, $(m\in\Z)$, be subspaces of $M$ such that
$\cdots \subset N_m\subset N_{m+1}\subset\cdots$ and
$\bigcap_{m\in\Z}N_m = \{\,0\,\}$. Then there exists a
nondegenerate subspace $M^\vee\subset M^*$
such that for any $v^\vee\in M^\vee$, there
exists an $m\in \Z$ such that
$\pairing{v^\vee}{N_m} = 0$.
\end{lemma}
\begin{proof}
Set $N = \bigcup_{m\in\Z}N_m$ and consider
\[
N^\vee_m = \{\,v^\vee\in N^* \,|\,\pairing{v^\vee}{N_m}
= 0\,\}\subset N^*.
\]
Take a complement $P$ of $N$ in $M$ so that $M =
N\oplus P$ and define 
\[
M^\vee = \biggl(\bigcup_{m\in\Z}N^\vee_m\biggr)\oplus
P^*\subset M^*.
\]
Then this $M^\vee$ has the desired properties.
\end{proof}

Let us recall that a series $A(z) =
\sum_{n\in\Z}A_nz^{-n-1}$ on $M$ is called a field
if for any $u\in M$, we have $A_nu = 0, (n\geq
n_0)$, for some $n_0\in\Z$. Then, a series
$A(z)$ is a field if and only if 
\[
\bigcup_{n_0\in\Z}\biggl(\bigcap_{n=n_0}^\infty\Ker{A_n}\biggr)
= M.
\]
Now we define the dual notion as follows: A series
$A(z)$ on $M$  is said to be
\textit{cotruncated}\, if for any nonzero
$u\in M$, we have
$u\notin\operatorname{Im}\,A_n, (n<n_0)$,
for some $n_0\in\Z$. Then, a series $A(z)$ is 
cotruncated if and only if 
\[
\bigcap_{n_0\in\Z}\biggl(\sum_{n= -\infty}^{n_0}
\operatorname{Im}\, A_n\biggr) = \{\,0\,\}.
\]
If $A(z)$ is cotruncated, then by Lemma
\ref{lemma:6.1.1}, there exists a nondegenerate
subspace $M^\vee$ such that, for any $v^\vee\in
M^\vee$, there exists $n_0$ satisfying
$\pairing{v^\vee}{A_nu} = 0,(n<n_0)$,
for all $u\in M$.

More generally, we prepare the following notion: We say
that series $A^{(1)}(z),\dots,A^{(\ell)}(z)$ are
\textit{admissible\,} if, for any $u\in M$, 
\[
\bigcap_{m\in\Z}\biggl(
\sum_{p_1+\cdots +p_\ell = -\infty}^m
\sum_{\sigma\in\mathfrak{S}_\ell}\C
A^{(\sigma(1))}_{p_1}\cdots A^{(\sigma(\ell))}_{p_\ell}u
\biggr) = \{\,0\,\},
\]
where $\mathfrak{S}_\ell$ is the symmetric
group acting on $\{1,\dots,\ell\}$.
\begin{remark}\label{remark:6.1.2}
Suppose given a direct sum decomposition
\[
M = \oplus_{\lambda\in\C}M_\lambda =
\bigoplus_{[\lambda_0]\in\C/\Z}\biggl(
\oplus_{n\in\Z}M_{\lambda_0+n}\biggr)
\]
such that 
$
A^{(i)}_n(M_\lambda)\subset M_{\lambda+h_i-n},
(h_i\in \Z)$,
for $A^{(1)}(z),\dots,A^{(\ell)}(z)$.
If the set
$
\{\, n\in\Z\,|\, M_{\lambda_0+n} \neq 0\,\}
$
is bounded below, then the series
$A^{(1)}(z),\dots,A^{(\ell)}(z)$
are admissible fields. More generally, if there is a
filtration
\[
\mathcal{F}_0M\subset \mathcal{F}_1M\subset\cdots,
\quad \bigcup_{k=0}^\infty \mathcal{F}_kM=M,
\]
such that
$
\mathcal{F}_kM = \oplus_{\lambda\in
\C}(\mathcal{F}_kM)\cap M_\lambda,
$
the set 
$
\{\,n\in\Z\,|\,(\mathcal{F}_kM)_{\lambda+n}\neq 0\,\}
$
is bounded below, and  
$
{A^{(i)}_n}^{-1}(\mathcal{F}_kM)\subset
\mathcal{F}_{k+r_i(k)}M
$,
then the series
$A^{(1)}(z),\dots,A^{(\ell)}(z)$ are admissible fields.
\end{remark}

Now, let $A^{(1)}(z),\dots,A^{(\ell)}(z)$ be admissible
series. Then, for any $u\in M$, there exists a
nondegenerate subspace $M^\vee_u\subset M^*$
such that for any $v^\vee\in M^\vee_u$ there
exists an integer
$m_0\in\Z$ satisfying 
\[
\pairing{v^\vee}{A^{(\sigma(1))}_{p_1}
\cdots A^{(\sigma(\ell))}_{p_\ell}u} = 0,\quad 
(p_1,\dots,p_\ell\in\Z, p_1+\cdots
+p_\ell<m_0, \sigma\in\mathfrak{S}_\ell).
\]
We call such a
$M^\vee_u\subset M^*$ a \textit{restricted dual
space} compatible with
$A^{(1)}(z),\dots,A^{(\ell)}(z)$ with respect to $u\in
M$.

\begin{lemma}\label{lemma:6.1.4}
Let $A(z)$ and $B(z)$ be admissible fields. Then for
any $u\in M$ and $v^\vee\in M^\vee_u$, 
\[
\pairing{v^\vee}{A(y)B(z)u} \in\C((y^{-1},z)),\quad
\pairing{v^\vee}{B(z)A(y)u} \in\C((y,z^{-1})).
\]
\end{lemma}
\begin{proof}
Since $B(z)$ is a field, $\pairing{v^\vee}{A(y)B(z)u}$
has only finitely many terms of negative degree in $z$.
On the other hand, we have
\[
\pairing{v^\vee}{A_pB_q} = 0\quad \text{if}\quad p+q\leq
m_0.
\]
If $p\leq m_0-q_0$, then we have
$\pairing{v^\vee}{A_pB_qu} = 0$ 
since either $q\geq q_0$ or $p+q\leq m_0$ holds.
Thus $\pairing{v^\vee}{A(y)B(z)u}
\in\C((y^{-1},z))$. Similarly we have
$\pairing{v^\vee}{B(z)A(y)u}\in\C((y,z^{-1}))$.
\end{proof}

\subsection{Borcherds
identity for local fields}\label{subsection:6.2}
Now, under the admissibility, the locality is
characterized as follows:
\begin{proposition}\label{proposition:6.1.5}
Let $A(z)$ and $B(z)$ be admissible fields. Then,
$A(z)$ and $B(z)$ are local if and only if, for any
$u\in M$ and $v^\vee\in M^\vee_u$,
$\pairing{v^\vee}{A(y)B(z)u}$ and
$\pairing{v^\vee}{B(z)A(y)u}$ are the expansions of the
same rational function of the form
\[
\frac{P(y,z)}{(y-z)^m},\quad
P(y,z)\in
\C[y,y^{-1},z,z^{-1}],
\]
to series convergent in $|y|>|z|$ and $|y|<|z|$
respectively.
\end{proposition}
\begin{proof}
Suppose $A(z)$ and $B(z)$ are local. Then,
\[
\pairing{v^\vee}{A(y)B(z)u}(y-z)^m =
\pairing{v^\vee}{B(z)A(y)u}(y-z)^m
\]
for some $m\gg 0$.
Since
\begin{equation}
\begin{aligned}\label{eqn:27}
&\pairing{v^\vee}{A(y)B(z)u}(y-z)^m\in\C((y^{-1},z)),\\
&\pairing{v^\vee}{B(z)A(y)u}(y-z)^m\in\C((y,z^{-1})),
\end{aligned}\end{equation}
they are equal to a Laurent polynomial
$
P(y,z)\in \C[y,y^{-1},z,z^{-1}].
$
By (\ref{eqn:27}), we have
\begin{equation}\label{eqn:28}
\pairing{v^\vee}{A(y)B(z)u}
=\rest{\frac{P(y,z)}{(y-z)^m}}{|y|>|z|},\quad
\pairing{v^\vee}{B(z)A(y)u}
=\rest{\frac{P(y,z)}{(y-z)^m}}{|y|<|z|}
\end{equation}
by Lemma \ref{lemma:1.1.1}.

Conversely, if (\ref{eqn:28}) holds for any $u\in M,
v^\vee\in M^\vee_u$, then
\[
\pairing{v^\vee}{\bigl(A(y)B(z)(y-z)^m - B(z)A(y)(y-z)^m
\bigr)u} = 0.
\]
Therefore, since $M^\vee_u$ is nondegenerate,
\[
\bigl(A(y)B(z)(y-z)^m - B(z)A(y)(y-z)^m\bigr)u = 0
\]
and the fields $A(z)$ and $B(z)$ are local by
definition.
\end{proof}

\begin{theorem}\label{theorem:6.2.3}
Let $A^{(1)}(z),\dots,A^{(\ell)}(z)$ be
admissible fields. If they are local, then for
any $u\in M$ and $v^\vee\in M^\vee_u$,
\[
\langle
v^\vee,A^{(1)}(z_1),\dots,A^{(\ell)}(z_\ell)u
\rangle
\]
and its permutations with respect to
$A^{(1)}(z),\dots,A^{(\ell)}(z)$
are the expansions of the same rational
functions of the form
\[
\frac{P(z_1,\dots,z_\ell)}{\underset{i<j}{\prod}(z_i-z_j)^{m_{ij}}},
\quad
P(z_1,\dots,z_\ell)\in\C[z_1,z_1^{-1},\dots,
z_\ell,z_\ell^{-1}],
\]
into the series of convergent in the
corresponding regions.
\end{theorem}
\begin{proof}
Consider $\langle v^\vee, A^{(1)}(z_1)\cdots
A^{(1)}(z_1)u\rangle$. It has only finitely
many terms of negative degree in $z_\ell$
because $A^{(\ell)}(z)$ is a field. On the
other hand, by Proposition
\ref{proposition:1.5.7}, there exists $m\in \N$
such that
\[
A^{(2)}_{p_2}\cdots A^{(\ell)}_{p_\ell}u = 0,
\quad 
(p_2 +\cdots+p_\ell\geq m),
\]
and by the admissibility, there exists $n\in
\N$ such that
\[
\langle v^\vee, A^{(1)}_{p_1}\cdots
A^{(\ell)}_{p_\ell}u\rangle = 0,\quad
(p_1 +\cdots+p_\ell< n).
\]
Therefore, $\langle v^\vee, A^{(1)}(z_1)\cdots
A^{(\ell)}(z_\ell)u\rangle$ has only finitely
many terms of positive degree in $z_1$. Similar
statements holds for its permutations.

Now, take sufficiently large $n_{ij}\in \N,
(i<j)$, and consider the series
\[
\langle v^\vee, A^{(1)}(z_1)\cdots
A^{(\ell)}(z_\ell)u\rangle
\prod_{i<j}(z_i-z_j)^{n_{i\ell}}
\]
and its permutations. Then by the locality,
they are equal to each other. In particular,
they are equal to the same Laurent polynomial
$P(z_1,\dots,z_\ell)$. Therefore, we have by
Lemma \ref{lemma:1.1.1}, for example,
\[
\langle v^\vee, A^{(1)}(z_1)\cdots
A^{(\ell)}(z_\ell)u\rangle =
\rest{\frac{P(z_1,\dots,z_\ell)}
{\prod_{i<j}(z_i-z_j)^{n_{i\ell}}}}
{|z_1|>\dots>|z_\ell|}.
\]
\end{proof}

In particular, if $A(z),B(z)$ and $C(z)$ are
admissible local fields, then the series
$\langle v^\vee,A(x)B(y)C(z)u\rangle$ is
analytically continued to its permutations 
as a rational function. Therefore, we are
allowed to prove the Borcherds identity for
local fields (Corollary \ref{corollary:n1})
in the following way\footnote{%
Special cases of such derivations are found in physics
literatures, e.g.,
\cite{BBS}, \cite{BS}, \cite{T}.}:
%
\[
\begin{split}
\sum_{i=0}^\infty&\binom{p}{i}(A(z)_{(r+i)}B(z))_{(p+q-i)}C(z)\\
&=\sum_{i=0}^\infty\binom{p}{i}
\oint_{C_z}\frac{dy}{2\pi \sqrt{-1}}
\oint_{C_y}\frac{dx}{2\pi \sqrt{-1}}
\,(x-y)^{r+i}(y-z)^{p+q-i}C(z)A(x)B(y)\\
&=\oint_{C_z}\frac{dy}{2\pi \sqrt{-1}}
\oint_{C_y}\frac{dx}{2\pi \sqrt{-1}}
\,(x-y)^r(y-z)^q(x-z)^{p}C(z)A(x)B(y)\\
&=\oint_{C_z}\frac{dy}{2\pi \sqrt{-1}}
\oint_{C_{y,z}}\frac{dx}{2\pi \sqrt{-1}}
\,(x-y)^r(y-z)^q(x-z)^{p}A(x)B(y)C(z)\\
&\quad\quad -\oint_{C_z}\frac{dy}{2\pi
\sqrt{-1}}
\oint_{C_{z}}\frac{dx}{2\pi \sqrt{-1}}
\,(x-y)^r(y-z)^q(x-z)^{p}B(y)A(x)C(z)\\
&=\oint_{C_z}\frac{dy}{2\pi \sqrt{-1}}
\oint_{C_{z}}\frac{dx}{2\pi \sqrt{-1}}
\,(x-y)^r(y-z)^q(x-z)^{p}A(x)B(y)C(z)\\
&\quad\quad -\oint_{C_z}\frac{dy}{2\pi
\sqrt{-1}}
\oint_{C_{z}}\frac{dx}{2\pi \sqrt{-1}}
\,(x-y)^r(y-z)^q(x-z)^{p}B(y)A(x)C(z)\\
&=\sum_{i=0}^\infty (-1)^i\binom{r}{i}
\oint_{C_z}\frac{dy}{2\pi \sqrt{-1}}
\oint_{C_{z}}\frac{dx}{2\pi \sqrt{-1}}
\,(x-y)^{p+r-i}(y-z)^{q+i}A(x)B(y)C(z)\\
&\quad - \sum_{i=0}^\infty (-1)^{r+i}\binom{r}{i}
\oint_{C_z}\frac{dy}{2\pi \sqrt{-1}}
\oint_{C_{z}}\frac{dx}{2\pi \sqrt{-1}}
\,(x-y)^{p+r-i}(y-z)^{q+i}B(y)A(x)C(z)\\
&=\sum_{i=0}^\infty (-1)^i\binom{r}{i}
(A(z)_{(p+r-i)}(B(z)_{(q+i)}C(z))-
(-1)^rB(z)_{(q+r-i)}(A(z)_{(p+i)}C(z)))
\end{split}
\]
where $C_y$ is a contour around $y$, $C_z$ is around
$z$, and $C_{y,z}$ is around both $y$ and $z$. Here we
have omitted writing $\langle v^\vee$, and
$u\rangle$. 

\subsection{Borcherds identity of vertex
algebra} Let $V$ be a vector space and suppose
given a map $Y:V\longrightarrow (\End
V)[[z,z^{_1}]]$. We further assume that the
series $Y(a,z)$ and $Y(b,z)$ are admissible
fields for any $a,b\in V$. For each
$a,b,c\in V$, we denote by $V^\vee_{abc}$ the
restricted dual space compatible with $Y(a,z)$ and
$Y(b,z)$ with respect $c$.

Consider the binary operations $a_{(n)}b$ defined by
 $Y(a,z)b = \sum_{n\in\Z}a_{(n)}b
z^{-n-1}$. Then by Proposition
\ref{proposition:6.1.5}, the locality
(\ref{eqn:20}) holds if and only if,  for any
$c\in V$ and $v^\vee \in V^\vee_{abc}$\,,
\begin{align*}
\pairing{v^\vee}{Y(a,y)Y(b,z)c}& =
\rest{\frac{Q(y,z)}{y^kz^\ell(y-z)^m}}{|y|>|z|},\\
\pairing{v^\vee}{Y(b,z)Y(a,y)c}&
=\rest{\frac{Q(y,z)}{y^kz^\ell(y-z)^m}}{|y|<|z|}
\end{align*}
holds for some polynomial $Q(y,z)\in \C[y,z]$.

Then, under the locality, the duality (\ref{eqn:21})
means that
\[
\begin{split}
\pairing{v^\vee}{Y(Y(a,x)b,z)c}(x+z)^p
& =
\rest{\pairing{v^\vee}{Y(a,x+z)Y(b,z)c}}{|x|<|z|}(x+z)^p\\
&=\rest{\frac{Q(x+z,z)}{(x+z)^kz^\ell
x^m}}{|x|<|z|} (x+z)^p
\end{split}
\]
for the polynomial $Q(y,z)$ as above.
Substituting 
$x = y-z$ and using Lemma \ref{lemma:1.1.1},
we rewrite this as
\[
\pairing{v^\vee}{Y(Y(a,y-z)b,z)c} = 
\rest{\frac{Q(y,z)}{y^kz^\ell(y-z)^m}}{|y-z|<|z|}.
\]

Thus we arrive at the following condition:

\noindent
\textbf{(R)} For any $a,b,c\in V$ and any
$v^\vee\in V^\vee_{abc}$, the series
\begin{align*}
\pairing{v^\vee}{Y(a,y)Y(b,z)c},&\quad |y|>|z|,\\
\pairing{v^\vee}{Y(b,z)Y(a,y)c},&\quad |y|<|z|,\\
\pairing{v^\vee}{Y(Y(a,y-z)b,z)c},&\quad
|y-z|<|z|
\end{align*}
are the expansions of the same rational function of
the form
\[
\frac{Q(y,z)}{y^kz^\ell(y-z)^m},
\quad Q(y,z)\in \C[y,z],
\]
into  series convergent 
in the respective regions.

Summarizing the consideration above, we have
(cf. [FHL, Proposition 3.4.1])
\begin{proposition}\label{proposition:6.3.1}
Let $V$ be a vector space and suppose given a
map $Y:V\longmapsto (\End V)[[z,z^{-1}]]$. If
$Y(a,z)$ and
$Y(b,z)$ are admissible fields for any
$a,b\in V$, then the axiom (B1) is equivalent to the
property (R).
\end{proposition}

\subsection*{A. Appendix--List of expansions
of $(x-y)^r(y-z)^q(x-z)^p$}\label{appendix}
Let $F(x,y,z) = (x-y)^r(y-z)^q(x-z)^p$. We will give
the list of power series expansions of $F(x,y,z)$ in
various regions.
\subsubsection*{A.1. The expansion in the region
$|y-z|>|x-y|$}

\begin{align*}
&F(x,y,z)=
\sum_{i=0}^\infty\binom{p}{i}(x-y)^{r+i}(y-z)^{p+q-i}&\quad\\
&=^*\sum_{i,j,k=0}^\infty(-1)^{j+k}\binom{p}{i}\binom{r+i}{j}
\binom{p+q-i}{k}x^{r+i-j}y^{p+q-i+j-k}z^k,&\quad
(|x|>|y|>|z|)\\
&=^*\sum_{i,j,k=0}^\infty(-1)^{r+i+j+k}\binom{p}{i}\binom{r+i}{j}
\binom{p+q-i}{k}x^{j}y^{p+q+r-j-k}z^k,&
\quad(|y|>|x|>|z|)\notag\\
&=\sum_{i,j,k=0}^\infty(-1)^{p+q+i+j+k}\binom{p}{i}\binom{r+i}{j}
\binom{p+q-i}{k}x^{r+i-j}y^{j+k}z^{p+q-i-k},&\quad
(|z|>|x|>|y|)\\
&=\sum_{i,j,k=0}^\infty(-1)^{j+k}\binom{p}{i}\binom{r+i}{j}
\binom{p+q-i}{k}x^{j}y^{r+i-j+k}z^{p+q-i-k},&\quad
(|z|>|y|>|x|)
\end{align*}
Here the expressions with $=^*$ do not make sense for
$p<0$.

\subsubsection*{A.2. The expansion in the region
$|x-z|>|y-z|$}
\begin{align*}
&F(x,y,z)=
\sum_{i=0}^\infty(-1)^{i}\binom{p}{i}(x-z)^{p+r-i}
(y-z)^{q+i}&\\
&=\sum_{i,j,k=0}^\infty(-1)^{i+j+k}
\binom{r}{i}\binom{p+r-i}{j}\binom{q+i}{k}
x^{p+r-i-j}y^{q+i-k}z^{j+k},&\quad
(|x|>|y|>|z|)\\
&=\sum_{i,j,k=0}^\infty(-1)^{p+r+i+j+k}
\binom{r}{i}\binom{p+r-i}{j}\binom{q+i}{k}
x^{p+r-i-j}y^{k}z^{j+q+i-k},&\quad
(|x|>|z|>|y|)\\
&=^{**}\sum_{i,j,k=0}^\infty(-1)^{i+j+k}
\binom{r}{i}\binom{p+r-i}{j}\binom{q+i}{k}
x^{j}y^{q+i-k}z^{p+r-i-j+k},&\quad
(|y|>|z|>|x|)\\
&=^{**}\sum_{i,j,k=0}^\infty(-1)^{p+q+r+i+j+k}
\binom{r}{i}\binom{p+r-i}{j}\binom{q+i}{k}
x^{j}y^{k}z^{p+q+r-j-k},&\quad
(|z|>|y|>|x|)
\end{align*}
Here the expressions with $=^{**}$ do not make sense for
$r<0$.

\subsubsection*{A.3. The expansion in the region
$|y-z|>|x-z|$}
\begin{align*}
&F(x,y,z)=\sum_{i=0}^\infty(-1)^{r+i}
\binom{r}{i}(y-z)^{q+r-i}
(x-z)^{p+i}&\\
&=
\sum_{i,j,k=0}^\infty(-1)^{r+i+j+k}
\binom{r}{i}\binom{q+r-i}{j}\binom{p+i}{k}
x^{p+i-k}y^{q+r-i-j}z^{j+k},&\quad
(|y|>|x|>|z|)\\
&=\sum_{i,j,k=0}^\infty(-1)^{p+r+j+k}
\binom{r}{i}\binom{q+r-i}{j}\binom{p+i}{k}
x^{k}y^{q+r-i-j}z^{p+i+j-k},&\quad
(|y|>|z|>|x|)\\
&=^{**}\sum_{i,j,k=0}^\infty(-1)^{r+i+j+k}
\binom{r}{i}\binom{q+r-i}{j}\binom{p+i}{k}
x^{p+i-k}y^{j}z^{p+r-i-j+k},&\quad
(|x|>|z|>|y|)\\
&=^{**}\sum_{i,j,k=0}^\infty(-1)^{p+q+i+j+k}
\binom{r}{i}\binom{q+r-i}{j}\binom{p+i}{k}
x^{k}y^{j}z^{p+q+r-j-k},&\quad
(|z|>|x|>|y|)
\end{align*}
Here the expressions with $=^{**}$ do not make sense for
$r<0$.

\end{document}